# Confrontation with the West and Long-Run Economic and Institutional Outcomes: Evidence from Iran[1]

Rok Spruk

**Abstract**

*This paper studies the long-run economic and institutional consequences of Iran's confrontation with the West, treating the 2006–2007 strategic shift as the onset of a sustained confrontation regime rather than a discrete sanctions episode. Using synthetic control and generalized synthetic control methods, I construct transparent counterfactuals for Iran's post-confrontation trajectory from a donor pool of countries with continuously normalized relations with the West. I find large, persistent losses in real GDP and GDP per capita, accompanied by sharp declines in foreign direct investment, trade integration, and non-oil exports. These economic effects coincide with substantial and durable deterioration in political stability, rule of law, and control of corruption. Magnitude calculations imply cumulative output losses comparable to civil-war settings, despite the absence of internal armed conflict. The results highlight confrontation as a deep and persistent economic and institutional shock, extending the literature beyond short-run sanctions effects to sustained geopolitical isolation.*

# 1 Introduction

Iran's modern economic history is often narrated as a sequence of discrete crises and geopolitical escalations. Yet what distinguishes Iran is not merely exposure to international sanctions (Farzanegan and Batmanghelidj 2023), but the persistence of a broader political state of sustained confrontation with the West, encompassing recurring sanctions, diplomatic isolation, financial and technological restrictions, elevated geopolitical risk, and repeated disruptions to external integration (Laudati and Pesaran 2023). This paper asks a simple but under-studied question: What is the long-run causal effect of sustained international confrontation on trajectories of economic growth and institutional quality? In particular, can confrontation, without large-scale domestic warfare, generate growth and institutional losses comparable to those associated with civil conflict and revolutionary rupture?

Answering this question matters for both political economy and international economics. A large literature quantifies the economic effects of sanctions, emphasizing trade, welfare, and macroeconomic outcomes (Haidar 2017, Gharehgozli 2017, Afesorgbor 2019, Farzanegan and Habibi 2025). Recent work has substantially improved measurement and identification of sanctions episodes and their effects, including structural-gravity approaches to trade and welfare and event-study designs for growth and transmission channels (Garfinkel et. al. 2020, Kelishomi and Nistico 2022). For instance, Felbermayr et. al. (2020, 2021) develop comprehensive sanctions data and quantify heterogeneous trade and welfare effects, with explicit attention to Iran as a high-stakes case (Felbermayr 2025). In addition, Afesorgbor (2019) distinguishes threatened from imposed sanctions and shows that the effects can differ sharply, underscoring the importance of expectations and anticipation. More recently, Gutmann et. al. (2023) document economically meaningful declines in growth and key components following sanctions using event-study methods. In terms of further example, Biglaiser and Lektzian (2011) show that sanctions reduce U.S. foreign direct investment, highlighting a capital-flow channel through which coercion can bite in the medium and long run.

While this literature is essential, it leaves open a broader and arguably more structural question. Sanctions are often modelled and measured as identifiable policy instruments. By contrast, confrontation is a deliberate geopolitical regime that can reshape incentives, expectations, and domestic political equilibria. In that regime, economic isolation and elevated risk can alter not only outcomes such as GDP, investment, and trade (Campos et. al. 2023, Fernandez-Villaverde et. al. 2024), but also the institutions that govern economic exchange and political accountability (Kastner 2007). Empirically, the distinction matters. While sanctions may be a time-bounded episode, the confrontation is a long-lasting state that affects the level and trajectory of development through endogenous institutional change.

This paper provides new evidence on this question using Iran as a comparative case study. I estimate the long-run effects of confrontation with the West (i.e. United States, European Union and Israel) on (i) economic outcomes such as real GDP, GDP per capita, non-oil exports, FDI, trade openness, inflation and exchange-rate distortions, and (ii) institutional outcomes, captured by political stability, rule of law, control of corruption, and an aggregate institutional capacity index. Our empirical strategy combines three complementary approaches. First, I implement the synthetic control method (Abadie et. al. 2015) to construct a transparent counterfactual trajectory for Iran from a donor pool of unaffected countries, estimating long-run gaps in levels and trajectories. Second, to address concerns about latent common shocks and to recover a full dynamic treatment-effect path with uncertainty, I implement generalized synthetic control using interactive fixed effects following Xu (2017). Third,

I conduct in-space placebo permutations and in-time placebo tests to assess whether the estimated post-confrontation divergences are unusually large relative to what the method generates under reassigned treatments.

The results are striking and internally coherent. Across methods, Iran experiences a persistent shortfall in real economic activity relative to its counterfactual. Both aggregate utput and output per capita diverge downward for nearly two decades, with gaps that deepen rather than gradually close. Capital inflows collapse and remain depressed, non-oil exports underperform persistently, and exchange-rate distortions accumulate consistently with a durable external constraint rather than transient macroeconomic instability. Translating the estimated log gaps into levels implies economically large losses in annual output relative to the counterfactual, and cumulative losses that are comparable in order of magnitude to those found in settings typically associated with internal conflict.

Institutions deteriorate in parallel. Under the geopolitical confrontation with the West, political stability weakens, rule of law erodes, corruption control declines, and aggregate institutional capacity falls substantially below the counterfactual. The institutional effects are not merely contemporaneous but accumulate over time, consistent with confrontation inducing a shift in the domestic governance equilibrium. Importantly, these patterns survive stringent placebo tests and remain qualitatively and quantitatively unchanged when moving from classical synthetic control to generalized synthetic control as magnitudes attenuate, but the sign, persistence, and economic significance remain.

Our findings connect directly to a growing literature emphasizing that geopolitical conditions are economically consequential even outside formal war and even when policy instruments are diffuse. Measures of geopolitical risk show persistent links to economic activity and investment; for example, the geopolitical risk index in Caldara and Iacoviello (2022) documents economically meaningful and persistent real effects of geopolitical risk shocks. More broadly, emerging work measures geopolitical barriers and political distance directly from events and narrative data, allowing researchers to study how hostile relations reshape globalization and economic allocation over long horizons.

This paper contributes to that frontier in two ways. Substantively, it shows that sustained confrontation can generate civil war-scale economic and institutional losses in peacetime, operating through a self-reinforcing interaction between external constraints and endogenous institutional deterioration. Methodologically, it demonstrates how combining synthetic control, generalized synthetic control, and layered placebo designs can deliver credible long-run counterfactual inference in settings where "treatment" is a geopolitical regime rather than a single policy act. The remainder of the paper proceeds as follows. Section 2 describes the historical and policy background. Section 3 discusses the data. Section 4 presents the results develops inference via in-space placebo permutations and in-time placebos alongside the discussion and policy implications. Section 5 concludes.

## 2      Historical and Policy Background

Following the 1979 Islamic Revolution, Iran entered a prolonged phase of political consolidation and state-led economic organization. The immediate post-revolutionary period and the Iran-Iraq War imposed severe economic costs and shaped early institutional arrangements, but these constraints did not lock Iran into permanent international isolation. From the late 1980s onward, successive governments pursued pragmatic economic policies, including reconstruction, gradual market-oriented reforms, and selective reintegration into global markets.

By the 1990s and early 2000s, Iran exhibited meaningful economic engagement with the outside world. Oil exports expanded, foreign firms, particularly in energy and infrastructure, operated in Iran, and macroeconomic performance tracked regional and global trends. While political relations with Western countries remained strained, economic interaction, investment, and trade were substantial (Maloney 2015). Importantly, this period establishes that Iran's economy and institutions were not on a deterministic path of isolation or decline prior to the mid-2000s (Salehi-Isfahani 2009). From an empirical perspective, this era provides a stable pre-treatment period in which Iran's economic and institutional trajectories evolve smoothly and are well matched by the synthetic counterfactual. There is no evidence of a pre-existing break or trend shift that would mechanically explain the post-2007 divergence documented later.

In the years immediately preceding 2006, Iran benefited from favorable external conditions. Global oil prices rose sharply, fiscal revenues increased, and the balance of payments strengthened. Inflationary pressures and structural inefficiencies persisted, but these were characteristic of many resource-rich middle-income economies and did not prevent growth, investment, or trade. Crucially, Iran did not face binding multilateral sanctions during this period. Access to international financial markets, though imperfect, was not fundamentally disrupted, oil exports were unconstrained, and foreign firms continued to operate in Iran. These conditions are inconsistent with an economy already subject to a comprehensive sanctions regime or to effective economic containment.

This matters for identification: if the post-2007 outcomes reflected delayed responses to earlier sanctions or constraints, we would expect to observe pre-2006 distortions in investment, trade, or institutional indicators. Our evidence does not show such patterns.

The mid-2000s represent a clear political break. Beginning in 2005–2006, Iran explicitly shifted its external strategy toward confrontation with the West. This shift was driven by domestic political decisions, not by external compulsion. Key elements of this shift include the resumption and acceleration of uranium enrichment activities, the rejection of negotiated compromise proposals, the curtailment of cooperation with international monitoring arrangements, and a rhetorical and strategic framing of relations with Western governments as adversarial. These choices were publicly articulated and domestically endorsed as matters of sovereignty and deterrence.

Importantly, these decisions precede the imposition of the most economically consequential sanctions. United Nations and comprehensive financial sanctions escalate later, particularly after 2010-2012. Thus, the confrontation regime begins before sanctions bind and cannot be interpreted as a response to economic punishment. Rather, sanctions follow as a consequence of Iran's chosen external posture. From a causal standpoint, this establishes 2006-2007 as the onset of the treatment, a regime shift in geopolitical orientation, not a policy shock imposed from abroad.

Although comprehensive sanctions were not yet in place, the strategic shift of 2006-2007 immediately altered expectations and risk assessments. Foreign investors reassessed long-horizon projects, financial institutions increased caution, and uncertainty regarding market access and political risk rose sharply. These responses do not require formal sanctions. Elevated geopolitical risk alone can induce precautionary behavior, delay investment, and alter domestic political incentives. As a result, the economic and institutional consequences of confrontation begin to materialize before sanctions operate through mechanical trade or financial restrictions. Empirically, this pattern is reflected in the gradual divergence of Iran's outcomes from the synthetic counterfactual shortly after 2006-2007, with gaps widening over time. The absence of an abrupt collapse at the treatment onset is consistent with confrontation operating as a persistent equilibrium shock, not a one-time disruption.

A potential concern is that the estimated effects merely capture the impact of sanctions imposed after 2010. Several features of the analysis argue against this interpretation. First, the treatment onset is dated conservatively to 2006-2007, before sanctions bind. Second, the divergence in outcomes begins earlier and evolves smoothly, rather than coinciding with discrete sanctions episodes. Third, the estimated effects persist and deepen even in periods when sanctions intensity fluctuates, indicating that the underlying driver is the confrontation regime rather than any single policy instrument.

Conceptually, this paper treats sanctions as one mechanism through which confrontation operates, not as the confrontation itself. The empirical estimates therefore capture the cumulative impact of sustained geopolitical hostility including elevated risk, isolation, and endogenous institutional responses rather than the marginal effect of a particular sanctions package. Therefore, the political and economic background clarifies the interpretation of the empirical results. The treatment captures Iran's own strategic choice to enter a sustained confrontation with the West, beginning in 2006-2007. The estimated economic and institutional effects should therefore be understood as the long-run consequences of that choice, encompassing but not limited to the later imposition of sanctions. By anchoring the treatment to this earlier political turning point, the analysis isolates the broader equilibrium effects of confrontation and avoids conflating them with the narrower effects of sanctions. This distinction is essential for understanding both the magnitude and the persistence of the results documented in the remainder of the paper.

## 3 Identification Strategy and Data

### 3.1 Identification Strategy

The empirical strategy builds on the synthetic control method, which is designed to estimate the causal effect of a large, non-recurrent intervention affecting a single treated unit at a known point in time (Abadie et. al. 2015, Abadie 2021, Botosaru and Ferman 2019, Ferman and Pinto 2021). The key challenge in such settings is to construct a credible counterfactual trajectory for the treated unit and estimate of how outcomes would have evolved in the absence of Iran's confrontation with the West. Let $Y_{i,t}$ denote the outcome of interest for unit $i$ at time $t$. and let unit $i = 1$ correspond to Iran. Treatment begins at time $T_0$, which we identify with the onset of confrontation in 2006. The observed outcome can be written as:

$$Y_{i,t} = Y_{i,t}^N + \tau_{i,t} D_{i,t}$$

where $Y_{i,t}^N$ is the potential outcome in the absence of confrontation, $D_{i,t}$ is an indicator equal to one for Iran in periods $t > T_0$, and $\tau_{i,t}$ is time-varying treatment effect. Our key quantity of interest is $\tau_{1,t} = Y_{1,t} - Y_{1,t}^N$ for $t \geq T_0$. Because $Y_{1,t}^N$ is unobserved after the confrontation, the synthetic control method approximates it using a weighted combination of untreated donor units:

$$Y_{1,t}^N = \sum_{j=2}^{J+1} w_j Y_{j,t}$$

where $w_j$ are non-negative and sum to one. The weights are chosen to minimize the distance between Iran and the synthetic control in the pre-confrontation period, typically with respect to the full pre-confrontation path, and when appropriate, additional predictors (Doudchenko and Imbens 2016). The identifying logic is transparent. If a convex combination of untreated countries reproduces Iran's pre-confrontation trajectory closely, then, absent treatment, it provides a credible approximation to Iran's post-confrontation counterfactual (Kaul et. al. 2022). Under this condition, systematic divergence between Iran and its synthetic control after $T_0$ can be attributed to the confrontation shock.

To make this more precise let $W = (w_2, \ldots w_{J+1})'$ denote the vector of donor weights, restricted to simplex $\mathcal{W} = \{W: w_j \geq 0, \sum_{j=2}^{J+1} w_j = 1\}$. Let $X_1$ be a $k \times 1$ vector of pre-confrontation characteristics for Iran, and let $X_0$ be the corresponding $k \times J$ matrix collecting the same predictors for donor units. Predictors may include the full pre-confrontation outcome path (or selected lags), as well as other covariates prior to the shock. The synthetic control weights are not chosen arbitrarily but through the solution to the following convex optimization:

$$W(V) = \underset{W \in \mathcal{W}}{\mathrm{argmin}} (X_1 - X_0 W)' V (X_1 - X_0 W)$$

Where $V$ is a positive semi-definite $k \times k$ weighing matrix that governs the relative importance assigned to matching each predictor. In the canonical implementation, $V$ is selected to optimize pre-confrontation fit, typically by minimizing the mean square prediction error (MSPE) of the outcome in the pre-confrontation period:

$$\hat{V} = \underset{V \in \mathcal{V}}{\mathrm{argmin}} \frac{1}{T_0 - 1} \sum_{t=1}^{T_0+1} \left( Y_{1,t} - \sum_{j=2}^{J+1} \widehat{w}_j(V) Y_{j,t} \right)^2$$

Where $\mathcal{V}$ restricts $V$ to be diagonal and non-negative in standard applications. Substituting $\widehat{V}$ yields the final weights $\widehat{W} = \widehat{W}(\widehat{V})$, and the estimated counterfactual is $\widehat{Y}_{1,t}^N = \sum_{j=2}^{J+1} \widehat{w}_j Y_{j,t}$. This formulation clarifies that synthetic control constructs a counterfactual by choosing the unique convex combination of donor units that best reproduces Iran's pre-confrontation outcome path vector, most importantly, the pre-confrontation outcome trajectory, thereby making any post-confrontation divergence attributable to the confrontation shock under the maintained assumptions.

This approach rests on three key assumptions. First, there is no anticipation of the confrontation shock affecting outcomes prior to $T_0$. This is assessed empirically through pre-treatment fit and in-time placebo tests. Second, the donor units are not themselves affected by the treatment, which motivates the restriction of the donor pool to countries with normalized relations with the West and no sustained confrontation regime. Third, the relationship between the treated unit and the donor pool is stable over time in the absence of treatment, so that the pre-treatment match remains informative for post-treatment counterfactual construction.

Unlike conventional difference-in-differences designs, synthetic control does not rely on parallel trends across a large number of treated and untreated units. Instead, it constructs parallelism by design for the treated unit itself. This feature is particularly important in the present context, where the treatment is a geopolitical regime shift affecting a single country and where untreated countries differ widely in levels and growth rates. Inference is conducted using placebo-based methods that assess whether the estimated post-treatment gaps for Iran are unusually large relative to gaps obtained by reassigning the treatment to donor units or to pre-treatment periods. This allows us to evaluate the significance of the estimated effects without relying on large-sample approximations that are inappropriate in single-treated-unit settings. Finally, to address concerns about latent common shocks and time-varying unobservables, the analysis complements classical synthetic control with generalized synthetic control based on interactive fixed effects.

### 3.2 Outcomes, measurement and sample structure

The empirical analysis tracks Iran's macroeconomic performance and external adjustment using a parsimonious set of outcomes that jointly capture real activity, integration with global markets, and real distortions. To capture macroeconomic performance and adjustment, we include the following economic outcomes: (i) real GDP (2012 $Geary-Khamis international dollar, log), real GDP per capita (2012 $Geary-Khamis international dollar, log), trade openness (i.e. log share of exports and imports relative to GDP), non-oil exports (merchandise trade share of GDP, log), net FDI inflows (as a share of GDP), inflation rate (log-normalized), the foreign exchange rate (log-normalized), military expenditure (percent of GDP), and a purchasing-power-parity (PPP) measure. The data on economic outcomes is from World Bank's Development Indicators and Feenstra et. al. (2015) whilst the data on military expenditure comes from SERPI. The baseline pre-confrontation period spans 1996–2006, and the post-confrontation period begins in 2007, consistent with the interpretation of 2006/07 as the onset of a confrontation regime rather than the onset of later, binding sanctions.

Two practical features of this outcome set are worth emphasizing. First, the measures cover both real-side outcomes (i.e. production and income) and external constraint outcomes (FDI, trade margins, exchange rate, PPP, inflation), which is critical when the hypothesized mechanism operates through isolation, risk premia, and restricted integration. Second, the log specification for scale variables provides a transparent mapping between estimated effects and economically interpretable proportional losses, while the normalized real series allow the analysis to focus on sustained distortions rather than short-run volatility.

### 3.3 Pre-confrontation outcome fit

A central requirement for credible synthetic control inference is that the synthetic counterfactual reproduces the treated unit's pre-treatment trajectory closely. The pre-confrontation path-imbalance diagnostics reported in Table 1 establish that this requirement is met for the economic outcomes. Across the two primary real-side outcomes, the pre-treatment fit is tight. The pre-confrontation RMSPE is 0.024 for log real GDP and 0.021 for log real GDP per capita, and the synthetic-control bias is below 1 percent for both series. The corresponding fit statistics are near unity ($R^2 \approx$ 0.97-0.98), implying that the synthetic counterfactual tracks Iran's pre-confrontation evolution closely year by year. These diagnostics are particularly informative because the identifying variation in this paper comes from a regime change that is expected to shift trajectories gradually without a strong pre-treatment match, post-treatment divergence could be attributed to pre-existing drift rather than to the confrontation shock.

Several outcomes exhibit large average control but low synthetic control bias when the counterfactual is defined naively as the unweighted mean of donor countries, most notably FDI inflows and certain nominal measures. This is not a weakness of the design. Rather, it highlights precisely why synthetic control is appropriate in this setting. The average donor country is not a credible proxy for Iran. By construction, Iran is unusual in its economic structure and exposure to geopolitical risk even prior to confrontation. The synthetic-control algorithm addresses this by assigning weights to donor countries to match the pre-treatment path; accordingly, the synthetic-control bias falls sharply (e.g., to 6 percent for FDI inflows and below 1 percent for the foreign exchange rate and PPP), demonstrating that the counterfactual is not driven by average differences but by a targeted reconstruction of Iran's own pre-confrontation trajectory. These path-imbalance diagnostics are therefore not ancillary but are a key part of the identification argument. They show that (i) there is no mechanical pre-confrontation divergence, and (ii) the counterfactual is constructed to replicate Iran's own pre-confrontation evolution rather than to compare Iran to a generic or incomparable set of countries.

**Table 1**: Pre-Confrontation Macroeconomic Outcomes' Path Imbalance, 1996-2006

| | Real GDP (log) | | Real GDP per capita (log) | | Trade Openness (log) | | Non-Oil Exports (log) | | FDI net inflows, % | | Inflation rate (log-normalized) | | Foreign Exchange Rate (log-normalized) | | Military Expenditure (% GDP) | | Purchasing Power Parity (PPP) | |
|---|---|---|---|---|---|---|---|---|---|---|---|---|---|---|---|---|---|---|
| RMSE | 0.024 | | 0.021 | | 0.14 | | 0.02 | | 0.11 | | 0.25 | | 0.22 | | 0.33 | | 0.19 | |
| Average Control Bias (%) | 3.8% | | 2% | | 14.3% | | 33.1% | | 170% | | 26% | | 67% | | 7% | | 82% | |
| Synthetic Control Bias (%) | <1% | | <1% | | 1.8% | | 1.9% | | 6% | | 3.7% | | <1% | | 2.7% | | <1% | |
| R2 | 0.97 | | 0.98 | | 0.19 | | 0.45 | | 0.39 | | 0.54 | | 0.46 | | 0.40 | | 0.78 | |
| | Real | Synthetic | Real | Synthetic | Real | Synthetic | Real | Synthetic | Real | Synthetic | Real | Synthetic | Real | Synthetic | Real | Synthetic | Real | Synthetic |
| 1996 | 27.26 | 27.26 | 9.31 | 9.30 | 3.56 | 3.64 | 0.17 | 0.15 | 0.02 | 0.18 | 3.40 | 3.47 | 7.69 | 7.88 | 2.24 | 2.49 | 6.19 | 6.16 |
| 1997 | 27.28 | 27.32 | 9.31 | 9.33 | 3.49 | 3.71 | 0.15 | 0.16 | 0.05 | 0.23 | 2.91 | 2.96 | 7.93 | 8.01 | 2.50 | 2.66 | 6.32 | 6.26 |
| 1998 | 27.30 | 27.30 | 9.31 | 9.32 | 3.38 | 3.70 | 0.11 | 0.15 | 0.02 | 0.16 | 2.94 | 3.14 | 8.07 | 8.48 | 2.63 | 2.38 | 6.40 | 6.72 |
| 1999 | 27.32 | 27.33 | 9.32 | 9.32 | 3.55 | 3.55 | 0.17 | 0.17 | 0.03 | 0.14 | 3.05 | 3.01 | 8.34 | 8.52 | 2.40 | 2.46 | 6.66 | 6.84 |
| 2000 | 27.37 | 27.38 | 9.36 | 9.37 | 3.72 | 3.61 | 0.23 | 0.19 | 0.04 | 0.14 | 2.74 | 2.71 | 8.65 | 8.66 | 2.30 | 2.28 | 6.86 | 7.01 |
| 2001 | 27.40 | 27.42 | 9.37 | 9.39 | 3.70 | 3.89 | 0.18 | 0.19 | 0.32 | 0.39 | 2.51 | 2.63 | 8.73 | 8.79 | 2.42 | 2.60 | 6.97 | 7.14 |
| 2002 | 27.48 | 27.46 | 9.44 | 9.42 | 3.87 | 3.87 | 0.20 | 0.20 | 0.36 | 0.22 | 2.73 | 2.64 | 8.84 | 8.78 | 2.17 | 2.56 | 7.21 | 7.19 |
| 2003 | 27.56 | 27.51 | 9.51 | 9.46 | 3.93 | 3.88 | 0.20 | 0.20 | 0.44 | 0.25 | 2.86 | 2.27 | 9.01 | 8.75 | 2.41 | 2.65 | 7.31 | 7.23 |
| 2004 | 27.60 | 27.56 | 9.53 | 9.52 | 3.94 | 3.99 | 0.20 | 0.22 | 0.51 | 0.33 | 2.76 | 2.35 | 9.06 | 8.77 | 2.83 | 2.38 | 7.50 | 7.28 |
| 2005 | 27.63 | 27.62 | 9.54 | 9.55 | 4.00 | 3.96 | 0.22 | 0.23 | 0.61 | 0.59 | 2.67 | 2.34 | 9.10 | 8.82 | 3.04 | 2.31 | 7.66 | 7.39 |
| 2006 | 27.68 | 27.68 | 9.57 | 9.60 | 3.97 | 4.00 | 0.25 | 0.24 | 0.79 | 0.73 | 2.40 | 2.31 | 9.12 | 8.81 | 3.32 | 2.15 | 7.76 | 7.48 |

### 3.4  Donor pool: definition, identification logic and exclusion principles

The construction of the donor pool is central to the identification strategy, as it defines the counterfactual world against which Iran's post-confrontation outcomes are evaluated. The guiding principle is to approximate how Iran's economic and institutional trajectories would have evolved in the absence of a sustained confrontation with the West, holding constant global economic conditions and broad development trends. Accordingly, the donor pool consists of countries that satisfy three criteria. First, donor countries maintain normalized political and economic relations with the West, in particular with the United States, the European Union, and Israel, throughout the sample period. This condition ensures that donor countries are not themselves subject to a comparable regime of geopolitical hostility, diplomatic isolation, or systematic restrictions on trade, finance, or technology access. Second, donor countries do not engage in sustained strategic confrontation with Western powers during the sample period, even if they experience episodic political tensions or short-lived disputes. Third, donor countries are free of large-scale internal armed conflict, state collapse, or prolonged macroeconomic instability that would make their pre-treatment dynamics fundamentally incomparable to Iran's.

This donor pool design reflects a conceptual distinction that is central to the paper. The object of interest is not the effect of a specific policy instrument, such as a particular sanctions package, but the effect of a confrontation regime as a persistent political state characterized by elevated geopolitical risk, adversarial relations, and constrained external integration. Including countries that experience similar forms of confrontation, even through different mechanisms, would mechanically contaminate the counterfactual and bias estimated effects toward zero. Excluding such countries is therefore a requirement for internal validity rather than a discretionary modeling choice. Therefore, the donor pool consists of twelve middle-income countries and emerging economies that meet these criteria.[2]

From an identification perspective, the donor pool can be interpreted as representing the set of countries that approximate salient global environment facing middle-income and emerging economies that remain integrated into the international economic system. The synthetic control algorithm then constructs, for each outcome, a convex combination of these countries that reproduces Iran's pre-confrontation trajectory. In this sense, the donor pool defines the relevant comparison universe, while the weights determine the specific counterfactual path.

The exclusion restrictions embedded in the donor pool deserve emphasis. Countries subject to comprehensive sanctions, persistent diplomatic isolation, or severe geopolitical hostility toward the West are excluded even if they share regional, cultural, or resource-based similarities with Iran. Likewise, countries experiencing civil war, state failure, or extreme macroeconomic deterioration are excluded to prevent the counterfactual from being driven by dynamics unrelated to confrontation. These exclusions are not outcome-driven and are imposed ex ante based on observable political and geopolitical criteria.

An important implication of this design is that the synthetic counterfactual does not represent a comparison between Iran and average countries, nor does it rely on regional proximity or cultural similarity. Instead, it represents a disciplined approximation of how Iran's economy and institutions would have evolved had it remained embedded in the global economic system rather than entering a confrontation

---

[2] Bulgaria, Egypt, Indonesia, Jordan, Malaysia, Morocco, Romania, Serbia, South Africa, Tunisia, Turkey, Vietnam

regime. The credibility of this approximation is evaluated empirically through pre-confrontation path fit, reported in Section 3.2, rather than assumed a priori.

Finally, it is worth noting that the donor pool is intentionally conservative. By restricting attention to countries with normalized relations with the West, the analysis likely understates the full impact of confrontation, since the counterfactual excludes countries that face milder forms of geopolitical friction or partial isolation. The fact that large and persistent effects nonetheless emerge strengthens the interpretation that the estimated post-2007 divergences reflect the causal consequences of Iran's strategic choice to confront the West.

*3.5    Composition of synthetic control groups*

The synthetic control group is constructed separately for each outcome. As a result, the identity and weights of donor countries can differ across outcomes, reflecting differences in which countries best replicate Iran's pre-treatment path for a given variable. Table 2 reports these weights. The key interpretive point is that weights should not be read as importance or structural similarity in a general sense; rather, they are the algorithm's solution to a constrained matching problem for each outcome.

Several regularities nonetheless emerge and strengthen the credibility of the counterfactual. First, the counterfactual for real-side outcomes is composed of a small number of countries with stable macroeconomic dynamics that closely match Iran's pre-treatment levels and growth profile. For example, the synthetic control for real GDP is primarily composed of Indonesia, Jordan, and Vietnam, with a smaller contribution from Turkey. This pattern is exactly what should be expected in a well-behaved SCM application. The synthetic counterfactual is not an average of many unrelated countries, but a disciplined and salient convex combination that reproduces Iran's pre-confrontation macroeconomic outcome paths.

Second, for integration and external-balance margins, weights shift toward countries whose pre-treatment openness and non-oil export dynamics resemble Iran's. The non-oil exports counterfactual draws weight from Bulgaria, Malaysia, Serbia, Morocco, and Vietnam, reflecting the fact that export composition and diversification dynamics are not well matched by a single "typical" comparator.

Third, for FDI inflows, the synthetic control places substantial weight on Serbia and Turkey, with smaller weights on Bulgaria and Indonesia. This concentration is informative, not problematic. By default, FDI series are volatile and sensitive to both global cycles and domestic policy regimes, so the algorithm naturally relies on the subset of donor countries that most closely reproduce Iran's pre-treatment FDI dynamics. Importantly, the pre-confrontation bias diagnostics show that this weighted combination substantially improves fit relative to the naïve average donor.

Fourth, for adjustment outcomes such as the foreign exchange rate and PPP, the algorithm selects donor combinations that match Iran's pre-treatment normalization and trend. The foreign exchange rate counterfactual, for instance, relies heavily on Vietnam and Indonesia, with smaller weight on Serbia. This reflects the fact that nominal series often embed idiosyncratic policy regimes and normalization choices; the credibility of the counterfactual therefore rests primarily on pre-treatment fit rather than on superficial regional similarity. Finally, the military expenditure counterfactual draws on a different set of donors, including Bulgaria, Malaysia, Morocco, Tunisia, and Jordan. This outcome is structurally distinct. It partly reflects policy choice and part security environment, and it is desirable that its counterfactual is not mechanically driven by the same donors as output or trade. Overall, the weight patterns are consistent with a coherent SCM design. The counterfactual composition varies by outcome in ways that align with the economic structure of each series, while the pre-treatment path-imbalance diagnostics confirm that these weights achieve the intended objective of reproducing Iran's pre-confrontation evolution.

**Table 2**: Composition of Synthetic Control Groups Across Macroeconomic Outcomes

|  | Real GDP | Real GDP per capita | Trade Openness | Non-Oil Exports | FDI net inflows | Inflation rate | Foreign Exchange Rate | Military Expenditure | Purchasing Power Parity |
|---|---|---|---|---|---|---|---|---|---|
| Bulgaria | 0 | 0.12 | 0 | 0.31 | 0.04 | 0 | 0 | 0.25 | 0 |
| Egypt | 0 | 0 | 0.14 | 0 | 0 | 0.15 | 0 | 0 | 0 |
| Indonesia | 0.39 | 0 | 0 | 0 | 0.04 | 0 | 0.29 | 0 | 0.91 |
| Jordan | 0.31 | 0 | 0 | 0 | 0 | 0 | 0 | 0.04 | 0 |
| Malaysia | 0 | 0.34 | 0 | 0.20 | 0 | 0 | 0 | 0.32 | 0 |
| Morocco | 0 | 0.11 | 0 | 0.07 | 0 | 0 | 0 | 0.23 | 0 |
| Romania | 0 | 0 | 0 | 0 | 0 | 0.04 | 0 | 0 | 0 |
| Serbia | 0 | 0.17 | 0.18 | 0.18 | 0.65 | 0.30 | 0.13 | 0 | 0.09 |
| South Africa | 0 | 0 | 0 | 0 | 0 | 0 | 0 | 0 | 0 |
| Tunisia | 0 | 0.26 | 0 | 0 | 0 | 0 | 0 | 0.16 | 0 |
| Turkey | 0.07 | 0 | 0.67 | 0 | 0.26 | 0.27 | 0 | 0 | 0 |
| Vietnam | 0.23 | 0 | 0 | 0.24 | 0 | 0.24 | 0.58 | 0 | 0 |

### 3.6  Advantages and limitations of the data design

This data design offers three main advantages. First, the long pre-treatment window and the breadth of outcomes allow the analysis to capture a confrontation regime whose effects are expected to be gradual and cumulative rather than instantaneous. Second, the outcome set spans both real activity and the external constraint, enabling a direct assessment of mechanisms consistent with isolation and elevated geopolitical risk. Third, the explicit path-imbalance diagnostics demonstrate that the synthetic counterfactual is empirically credible before treatment, a prerequisite for interpreting post-treatment divergence as causal.

At the same time, the design has limitations that discipline interpretation. Some series, particularly FDI and adjustment measures, are intrinsically volatile and may be measured with error or affected by definitional changes over time. This is addressed through normalization and through inference procedures that emphasize the distribution of placebo gaps rather than relying solely on conventional standard errors. In addition, the donor pool is necessarily constrained by the requirement of non-confrontation and normalized relations with the West, which improves internal validity but may limit external validity to settings with comparable geopolitical regimes and economic structures. Finally, as in any macroeconomic comparative case study, unobserved shocks correlated with the confrontation regime cannot be ruled out mechanically. Therefore, the empirical strategy triangulates across classical synthetic control, generalized synthetic control, and in-space and in-time placebo designs.

*3.7 Institutional outcomes: data, pre-confrontation fit, and donor composition*

Institutional outcomes are drawn from the Worldwide Governance Indicators (WGI), as developed by Kaufmann et. al. (2008). We focus on political stability and absence of violence, rule of law, and control of corruption, together with an aggregate institutional capacity index constructed as the first principal component of these measures. The WGI indicators are particularly well suited for the present analysis because they are explicitly designed to capture persistent institutional characteristics rather than short-run policy fluctuations, and because they are available consistently for a long pre-treatment window.

As with the macroeconomic outcomes, identification relies on the ability of the synthetic control to reproduce Iran's pre-confrontation institutional trajectory. Table 3 shows that this condition is satisfied. For all institutional measures, the synthetic-control bias is small, below 4 percent for political stability and rule of law and below 1 percent for the aggregate institutional component and corruption control, despite substantial bias in naïve average-control comparisons. Pre-treatment fit statistics are economically meaningful, with $R^2$ values ranging from 0.57 to 0.84, which is notably strong given the well-known measurement noise and bounded nature of institutional indicators. These diagnostics are particularly important in the institutional context. Institutional series are smoother and less volatile than macroeconomic outcomes, but they are also subject to cross-country level differences that cannot be addressed by simple demeaning or differencing. The fact that the synthetic control reproduces Iran's pre-2007 institutional path closely, year by year and across multiple dimensions, supports interpreting post-confrontation divergence as causal rather than as an artifact of pre-existing institutional drift.

**Table 3**: Pre-Confrontation Institutional Outcomes' Path Imbalance, 1996-2006

|  | Principal Component | | Political Stability | | Rule of Law | | Control of Corruption | |
|---|---|---|---|---|---|---|---|---|
| RMSE | 0.574 | | 0.07 | | 0.18 | | 0.03 | |
| Average Control Bias (%) | 104% | | 57% | | 74% | | 25% | |
| Synthetic Control Bias (%) | <1% | | 3.3% | | 3.5% | | <1% | |
| R2 | 0.65 | | 0.71 | | 0.57 | | 0.84 | |
|  | Real | Synthetic | Real | Synthetic | Real | Synthetic | Real | Synthetic |
| 1996 | -2.93 | -2.25 | -0.32 | -0.45 | -0.94 | -0.67 | -0.48 | -0.47 |
| 1997 | -2.88 | -2.71 | -0.51 | -0.54 | -0.81 | -0.71 | -0.47 | -0.45 |
| 1998 | -2.83 | -3.18 | -0.69 | -0.62 | -0.68 | -0.75 | -0.46 | -0.44 |
| 1999 | -2.73 | -2.81 | -0.71 | -0.69 | -0.60 | -0.73 | -0.43 | -0.45 |
| 2000 | -2.64 | -2.43 | -0.73 | -0.76 | -0.52 | -0.71 | -0.40 | -0.46 |
| 2001 | -2.68 | -2.71 | -0.75 | -0.79 | -0.67 | -0.77 | -0.31 | -0.34 |
| 2002 | -2.73 | -2.99 | -0.77 | -0.82 | -0.82 | -0.83 | -0.22 | -0.22 |
| 2003 | -2.64 | -3.16 | -0.78 | -0.78 | -0.68 | -0.79 | -0.31 | -0.30 |
| 2004 | -2.64 | -2.73 | -0.76 | -0.81 | -0.67 | -0.69 | -0.36 | -0.38 |
| 2005 | -3.11 | -2.50 | -0.78 | -0.64 | -0.87 | -0.66 | -0.48 | -0.41 |
| 2006 | -3.60 | -2.10 | -1.04 | -0.95 | -0.99 | -0.60 | -0.48 | -0.49 |

The donor pool for institutional outcomes is identical in principle to that used for macroeconomic outcomes and follows the same exclusion logic: donor countries maintain normalized relations with the United States, the European Union, and Israel, and do not engage in sustained geopolitical confrontation during the sample period. This restriction is especially important for institutions. Including donor countries subject to comparable external hostility or isolation would contaminate the counterfactual by embedding institutional deterioration driven by similar geopolitical pressures, thereby biasing estimated effects toward zero. Table 4 reports the composition of the synthetic control across institutional outcomes. As in the macroeconomic case, weights differ by outcome, reflecting the fact that no single donor country can replicate Iran's pre-treatment institutional profile along all dimensions simultaneously. Nonetheless, the pattern of non-zero weights is substantively coherent.

For the aggregate institutional principal component, the synthetic control assigns substantial weight to Indonesia, with smaller contributions from Egypt and Serbia. Indonesia's prominence is intuitive: it exhibits a long-run institutional trajectory characterized by gradual reform, episodic political stress, and post-authoritarian consolidation without sustained external confrontation. This makes it a natural comparator for Iran's pre-2007 institutional dynamics, even though the political systems differ. Egypt contributes weight because its pre-treatment political stability and corruption-control paths closely track Iran's levels and trends during the late 1990s and early 2000s, while remaining embedded in a non-confrontational external environment.

**Table 4**: Composition of Synthetic Control Groups Across Macroeconomic Outcomes

|  | Real GDP | Real GDP per capita | Trade Openness | Non-Oil Exports | FDI net inflows | Inflation rate | Foreign Exchange Rate | Military Expenditure | Purchasing Power Parity |
|---|---|---|---|---|---|---|---|---|---|
| Bulgaria | 0 | 0.12 | 0 | 0.31 | 0.04 | 0 | 0 | 0.25 | 0 |
| Egypt | 0 | 0 | 0.14 | 0 | 0 | 0.15 | 0 | 0 | 0 |
| Indonesia | 0.39 | 0 | 0 | 0 | 0.04 | 0 | 0.29 | 0 | 0.91 |
| Jordan | 0.31 | 0 | 0 | 0 | 0 | 0 | 0 | 0.04 | 0 |
| Malaysia | 0 | 0.34 | 0 | 0.20 | 0 | 0 | 0 | 0.32 | 0 |
| Morocco | 0 | 0.11 | 0 | 0.07 | 0 | 0 | 0 | 0.23 | 0 |
| Romania | 0 | 0 | 0 | 0 | 0 | 0.04 | 0 | 0 | 0 |
| Serbia | 0 | 0.17 | 0.18 | 0.18 | 0.65 | 0.30 | 0.13 | 0 | 0.09 |
| South Africa | 0 | 0 | 0 | 0 | 0 | 0 | 0 | 0 | 0 |
| Tunisia | 0 | 0.26 | 0 | 0 | 0 | 0 | 0 | 0.16 | 0 |
| Turkey | 0.07 | 0 | 0.67 | 0 | 0.26 | 0.27 | 0 | 0 | 0 |
| Vietnam | 0.23 | 0 | 0 | 0.24 | 0 | 0.24 | 0.58 | 0 | 0 |

Political stability synthetic trajectory draws primarily on Jordan and Indonesia. Both countries exhibit pre-treatment stability dynamics that mirror Iran's combination of regime durability and underlying political risk, without experiencing systematic external isolation. Jordan's weight is particularly informative. It reflects similarity in internal political constraints and regional exposure, while differing sharply from Iran in its external alignment, precisely the contrast required for a credible counterfactual. For rule of law, weights concentrate on Indonesia, Serbia, and Vietnam. These countries share a pattern of gradual legal and institutional evolution from weaker initial conditions, with incremental improvements and setbacks rather than sharp breaks. Their inclusion reflects similarity in institutional trajectories, not in political ideology or regional location, which is the appropriate criterion in this setting. Control of corruption synthetic peer draws weight from Egypt, Vietnam, and Tunisia. This composition reflects the fact that corruption indicators are especially sensitive to administrative practices and enforcement capacity rather than to regime type per se. The synthetic control therefore relies on countries whose pre-treatment corruption dynamics align closely with Iran's, while remaining outside a confrontation regime.

Two features of the institutional donor composition deserve emphasis. First, the synthetic control never relies on a single donor country; weights are distributed across a small set of countries whose combined trajectory reproduces Iran's pre-confrontation path. Second, the donors with non-zero weights are not selected because share similarites Iran in a broad sense, but because they replicate Iran's institutional evolution prior to confrontation while remaining embedded in normalized international relations. This distinction is essential as the synthetic counterfactual is not a political twin, but a reasonable approximation of institutional trajectory under non-confrontation. Finally, it is worth noting that the institutional donor composition reinforces the conservative nature of the design. By restricting attention to countries without sustained geopolitical hostility toward the West, the counterfactual likely understates the institutional consequences of confrontation. The fact that large and persistent post-2007 institutional gaps nonetheless emerge strengthens the conclusion that confrontation operates as a deep and durable force shaping institutional equilibria.

## 4 Results

### 4.1 Design setup and estimand

We estimate the causal effect of Iran's post-2006/2007 confrontation regime characterized by sanctions escalation and sustained geopolitical isolation on macroeconomic performance and institutional quality using synthetic control method (Abadie et. al. 2015). The baseline estimand is the average post-confrontation gap between Iran's realized outcome path and the outcome path of a weighted combination of non-treated countries that best reproduces Iran's *pre-confrontation* trajectory.

Formally, for each outcome $Y_t$, the baseline SCM constructs weights $w_j \geq 0$, such that the synthetic series $Y_t^{SC} = \sum_j w_j Y_{j,t}$ matches Iran's pre-treatment path as closely as possible. The treatment effect is then $\tau_t = Y_t^{IRN} - Y_t^{SC}$ for $t > T_0$, and we report (i) the average post-treatment effect and (ii) the end-of-sample effect. This design is deliberately conservative. It attributes to "confrontation" only the component of Iran's post-2006 divergence that is not already implicit in its pre-2006 dynamics and common global shocks.

### 4.2 Baseline results

#### 4.2.1 Economic outcomes

This section reports the baseline synthetic control estimates of the long-run economic effects of Iran's confrontation with the West. The confrontation with the West is understood as the onset and persistence of intensified geopolitical conflict, sanctions, and international isolation beginning in the mid-2000s, which jointly altered Iran's external economic environment. The analysis compares Iran's realized post-shock trajectory to that of a synthetic counterfactual constructed to closely match Iran's pre-confrontation economic dynamics. All estimates are reported as post-shock gaps between Iran and its synthetic control. We present average post-shock effects to summarize long-run divergence and end-of-sample effects to capture the cumulative extent of the shock over time.

Table 5 reports the estimated outcome gaps between Iran and its synthetic peers, focussing on the average post-confrontation gap. The most striking result concerns aggregate output. Column (1) reports the estimated aggregate output effects of the confrontation regime. Following the confrontation shock, Iran's real GDP diverges sharply and persistently from its counterfactual path. The estimated average post-shock effect on log real GDP is -0.272 (standard error = 0.024), implying that Iran's real GDP is approximately 24 percent lower than it would have been absent the confrontation, on average over the post-shock period. The gap widens steadily over time, reaching an end-of-sample divergence of -0.454 log points, consistent with a cumulative and compounding effect rather than a temporary disruption.

Two features of this result deserve emphasis. First, the magnitude of the estimated gap is large relative to typical macroeconomic shocks. A persistent 20-25 percent output shortfall is well outside the range of normal business-cycle fluctuations and instead indicative of a structural shift in the long-run growth path. The temporal pattern of the gap, small initially, then widening over time, is consistent with confrontation operating through channels such as reduced investment, lower productivity growth, and sustained external constraints, rather than through a one-time output collapse. Second, the divergence reflects a relative effect: Iran does not merely experience low growth in absolute terms, but systematically underperforms a counterfactual path that is anchored in its own pre-shock dynamics and exposed to the same global economic environment. The result therefore captures the incremental cost of confrontation with the West, rather than global downturns or region-wide shocks.

The output losses translate directly into lower living standards. Column (2) reports the estimated per capita GDP gaps between Iran and its synthetic non-confrontation peer. For log real GDP per capita, the estimated average post-confrontation gap is -0.158 (standard error = 0.022), corresponding to an average 15 percent decline in income per person relative to the counterfactual. The end-of-sample gap reaches -0.224 log points, reinforcing the conclusion that the confrontation shock is associated with a persistent deterioration in living standards rather than a short-lived income loss. Importantly, the gap in GDP per capita closely mirrors that in aggregate GDP, suggesting that population dynamics do not drive the result. Instead, the decline reflects weaker income generation per individual, consistent with reduced productivity growth and diminished access to global markets and capital.

The onset of the confrontation with the West is associated with a marked reduction in Iran's integration into the global economy. As reported in column (3), the estimated average post-shock effect on trade openness (log) is -0.250, implying a sizable and persistent contraction in the ratio of trade to GDP relative to the counterfactual. Although the end-of-sample effect is smaller in magnitude, the negative gap remains throughout the post-shock period, indicating that Iran does not revert to its pre-confrontation degree of openness. This result is economically intuitive. Confrontation with the West operates directly through trade restrictions, financial sanctions, compliance costs, and heightened uncertainty, all of which raise the effective cost of cross-border transactions. The sustained decline in trade openness suggests that these frictions do not dissipate over time, but instead induce a lasting reorientation away from global markets.

The contraction in external integration is particularly consequential for non-oil exports. Column (4) reports the estimated average post-shock effect on non-oil exports (log) of approximately -0.072, corresponding to a persistent drop relative to the counterfactual path. While smaller in magnitude than the aggregate output or trade-openness effects, this result is economically meaningful: non-oil exports are precisely the margin along which diversification and long-run resilience are expected to occur. The decline in non-oil exports indicates that confrontation constrains not only aggregate trade volumes but also the composition of exports, limiting the ability of firms to enter and expand in non-hydrocarbon sectors. This mechanism is consistent with higher fixed costs of exporting, reduced access to foreign inputs and finance, and weaker participation in global value chains.

Foreign direct investment responds particularly strongly to the confrontation shock. Column (5) reports the key post-confrontation parameter, the average effect and end-of-sample effect. The estimated average post-confrontation effect on FDI net inflows, as percent of GDP is -1.487 (standard error = 0.140), with an end-of-sample effect of -1.687. These estimates imply a large and persistent collapse in foreign investment relative to the counterfactual. FDI is forward-looking and highly sensitive to geopolitical risk, policy uncertainty, and expectations about long-run market access. The magnitude and persistence of the FDI decline therefore suggest that confrontation fundamentally alters investors' beliefs about Iran's future integration into the global economy. In this sense, the FDI result is best interpreted not as a short-term reaction to sanctions, but as evidence of a durable increase in perceived country risk.

The confrontation policy regime is also associated with severe deterioration in macroeconomic stability. For the inflation rate (log-normalized), the estimated average post-confrontation gap, reported in column (6), is +1.013, indicating substantially higher inflation relative to the counterfactual path. For the log-normalized foreign exchange rate (column 7), the estimated average effect is +1.453, with an end-of-sample effect exceeding +3.4 log points, reflecting prolonged and cumulative currency depreciation. While the precise quantitative interpretation depends on the normalization of these variables, the direction and persistence of the effects are unambiguous. Confrontation with the West appears to exert sustained pressure on prices and the currency, consistent with restricted access to foreign exchange, reduced capital inflows, and reliance on monetary accommodation under external constraints. Crucially, inflation and depreciation move together with output losses, reduced trade, and collapsing FDI, forming a coherent macroeconomic pattern rather than isolated nominal disturbances. The purchasing power parity (PPP) measure shows a large positive divergence relative to the counterfactual, as reported in column (9). This result must be interpreted cautiously, as PPP measures are mechanically linked to inflation, exchange rates, and price-level adjustments. Rather than serving as an independent outcome, the PPP result is best viewed as complementary evidence of substantial relative-price distortions induced by prolonged isolation.

Not all economic outcomes respond uniformly to the confrontation shock. For military expenditure as a share of GDP, as indicated in column (8), the estimated average post-shock effect is statistically weak, and the end-of-sample effect is negative.

This absence of a robust effect is informative: it suggests that the observed economic deterioration is not mechanically driven by a simple reallocation of resources toward military spending, but instead reflects broader structural and external constraints. Figure 1 presents the economic effects of Iran's confrontation with the West for the full vector of outcomes.

*4.2.2. Interpreting magnitudes: annual losses, cumulative effects, and limits*

The estimated GDP gap of approximately 24 percent is economically very large, but its interpretation requires care. At current levels of economic activity, this gap corresponds to hundreds of billions of dollars per year in purchasing-power-parity terms and tens of billions of dollars per year at market exchange rates. These figures are accounting translations of the estimated log gap, not welfare estimates, and should not be interpreted as precise measures of social loss. Because the divergence persists over nearly two decades, the losses also accumulate over time. A simple back-of-the-envelope calculation suggests that, if Iran's counterfactual GDP absent confrontation would have averaged roughly USD 400 billion per year at market exchange rates, then a sustained 24 percent shortfall over the 2007-2024 period implies cumulative forgone output on the order of USD 1.5-2 trillion. This calculation is illustrative rather than exact, but it conveys the scale of the long-run divergence.

Comparatively, an output gap of this magnitude separates countries that began from broadly similar income levels but experienced sharply different development trajectories, such as South Korea versus Argentina, Spain versus Greece, or Poland versus Ukraine. The confrontation shock therefore appears to have shifted Iran onto a fundamentally different long-run economic path. At the same time, several caveats are essential. The estimates do not isolate individual policy instruments, do not measure welfare or distributional effects, and do not imply that all economic outcomes are driven solely by confrontation. Rather, they quantify the aggregate long-run divergence associated with sustained geopolitical isolation, conditional on Iran's pre-shock trajectory and global economic conditions.

In sum, the baseline results indicate that Iran's confrontation with the West is associated with a large, persistent, and internally coherent deterioration in economic performance. Real GDP and GDP per capita decline sharply relative to a credible counterfactual; trade openness and non-oil exports contract; foreign direct investment collapses; and nominal stability deteriorates through sustained inflation and currency depreciation. The magnitude, persistence, and coherence of these effects point to a long-run regime shift rather than a temporary shock, with economic costs comparable in scale to major structural disruptions
.

**Table 5**: Baseline Economic Effects of Iran's Confrontation with the West: Synthetic Control Estimates

|  | Real GDP (log) | Real GDP per capita (log) | Trade Openness (log) | Non-Oil Exports (log) | FDI net inflows, % | Inflation rate (log-normalized) | Foreign Exchange Rate (log-normalized) | Military Expenditure (% GDP) | Purchasing Power Parity (PPP) |
|---|---|---|---|---|---|---|---|---|---|
|  | (1) | (2) | (3) | (4) | (5) | (6) | (7) | (8) | (9) |
| Average Effect | -0.272*** | -0.158*** | -0.250*** | -0.072*** | -1.487*** | +1.013 | +1.453 | +0.369 | 1.453 |
|  | (0.024 | (0.022) | (0.141) | (0.021) | (0.140) | (0.252) | (0.191) | (0.454) | (0.191) |
| 95% Confidence Intervals | (-0.283, -0.260) | (-0.169, 0.147) | (-.320, -.179) | (-0.029, -0.012) | (-1.557, -1.417) | (0.887, 1.138) | (1.358, 1.547) | (-0.142, 0.593) | (1.358, 1.547) |
| End-of-sample effect | -0.454 | -0.224 | -0.132 | -0.075 | -1.687 | +0.929 | +3.459 | -0.262 | 3.459 |
| ATE p-value | 0.000 | 0.000 | 0.001 | 0.001 | 0.000 | 0.000 | 0.000 | 0.417 | 0.000 |

Notes: This table reports baseline synthetic control estimates of the economic effects of Iran's confrontation with the West. The treatment is dated to 2006/2007, corresponding to the onset of a sustained confrontation regime. Each column reports results for a separate outcome. Outcomes in columns (1) through (4) are in logarithms; column (5) reports net FDI inflows as a percentage of GDP; columns (6) and (7) are log-normalized inflation and exchange-rate indices; column (8) reports military expenditure as a percentage of GDP; column (9) reports a log-normalized purchasing power parity (PPP) measure. The Average Effect is the mean post-confrontation gap between Iran and its synthetic control. The end-of-sample effect reports the estimated gap in the final year of the sample. Parentheses report standard deviations of the post-treatment gaps. 95% confidence intervals are constructed from the empirical distribution of in-space placebo gaps obtained by reassigning the treatment to donor countries. ATE p-values report the fraction of placebo effects that are at least as large in absolute value as the estimated effect for Iran. Asterisks denote statistically significant effects at 10% (*), 5% (**), and 1% (***), respectively.

**Figure 1**: Economic Effects of Iran's Confrontation with the West: Actual vs. Synthetic Control, 1996-2024

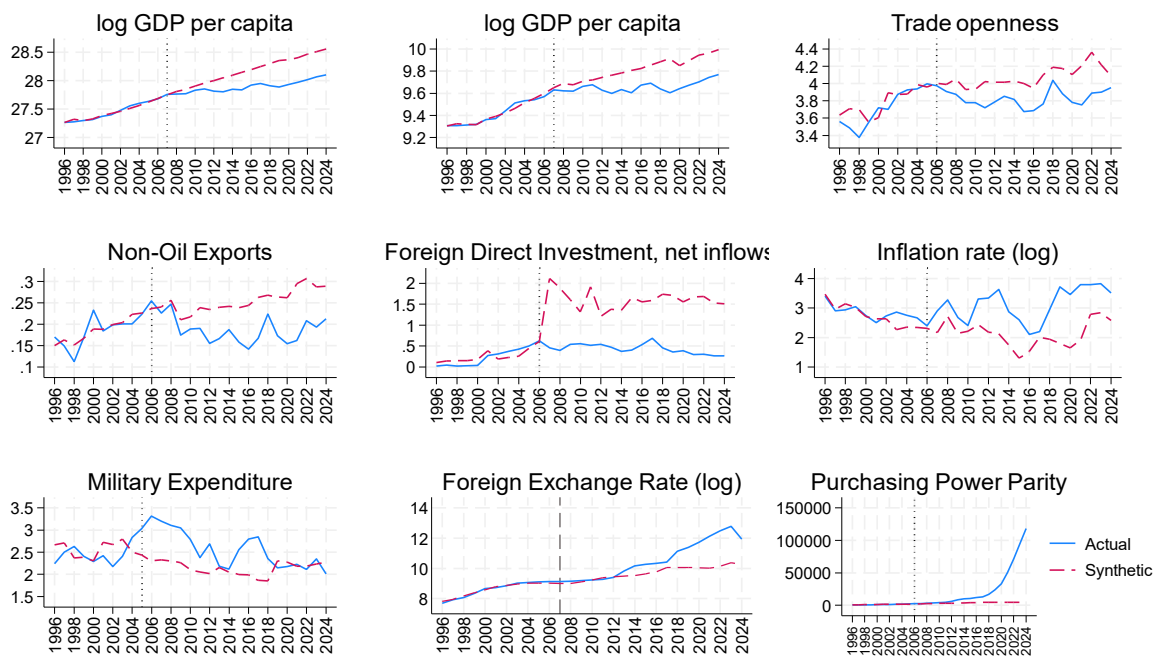

### 4.2.2 Institutional outcomes

Table 6 reports the effects of the confrontation with the West on institutional outcomes. The institutional effects associated with Iran's confrontation with the West are not only statistically significant and persistent, but also large in a way that is substantively meaningful in the cross-country distribution of governance outcomes. To assess their economic and institutional significance, this subsection interprets the estimated post-confrontation gaps relative to well-understood benchmarks in comparative institutional data.

**Table 6**: Institutional Effects of Iran's Confrontation with the West: Synthetic Control Estimates

|  | Principal Component | Political Stability | Rule of Law | Control of Corruption |
|---|---|---|---|---|
|  | (1) | (2) | (3) | (4) |
| Average Effect | -3.623*** | -0.889*** | -0.615*** | -0.459*** |
|  | (0.574) | (0.072) | (0.181) | (0.030) |
| 95% Confidence Intervals | (-3.908, 3.337) | (-0.925, 0.854) | (-0.705, -0.524) | (-0.473, -0.444) |
| End-of-sample effect | -6.142 | -1.394 | -1.058 | -0.712 |
| ATE p-value | 0.000 | 0.000 | 0.000 | 0.000 |

Notes: This table reports baseline synthetic control estimates of the institutional effects of Iran's confrontation with the West. The treatment is dated to 2006/2007, corresponding to the onset of a sustained confrontation regime. Each column reports results for a separate institutional outcome drawn from the Worldwide Governance Indicators (WGI). Column (1) reports the first principal component of institutional quality; columns (2) through (4) report Political Stability and Absence of Violence, Rule of Law, and Control of Corruption, respectively. The Average Effect is the mean post-confrontation gap between Iran and its synthetic control. Parentheses report standard deviations of the post-treatment gaps. 95% confidence intervals are constructed from the empirical distribution of in-space placebo gaps obtained by reassigning the treatment to donor countries. ATE p-values report the fraction of placebo effects that are at least as large in absolute value as the estimated effect for Iran. Asterisks denote statistically significant effects at 10% (*), 5% (**), and 1% (***), respectively.

In column (1), the estimated average post-confrontation effect on the institutional principal component is -3.62 standard deviations, with the end-of-sample gap exceeding -6.1 standard deviations. By the standards of cross-country institutional measurement, these magnitudes are exceptional. In most global governance datasets, differences on the order of one standard deviation typically separate broad institutional regimes, such as states with effective bureaucratic capacity and credible legal constraints from states characterized by weak enforcement and high discretion. A divergence of three to six standard deviations therefore implies not marginal institutional underperformance, but a fundamental repositioning of the state within the global institutional distribution. By the end of the sample period, Iran's institutional trajectory lies far outside the range of variation associated with gradual reform, policy experimentation, or cyclical political change. Instead, the magnitude of the gap is comparable to that observed between countries with relatively coherent institutional frameworks and those experiencing chronic governance failure. The temporal profile of the divergence is also informative. Rather than exhibiting an abrupt collapse, the principal component declines gradually but persistently following the confrontation shock, indicating cumulative institutional deterioration rather than a one-time governance disruption.

For political stability and absence of violence (column 2), the estimated average post-confrontation gap is -0.89 standard deviation, increasing to approximately -1.39 deviation by the end of the sample. In the cross-country distribution of political stability indicators, a gap of this magnitude typically separates countries characterized by relatively predictable political environments from those in which political violence, repression, or severe instability is a persistent risk. Interpreted in this way, the estimated gap reflects a qualitative change in the predictability of the political environment, rather than a marginal increase in episodic unrest. The persistence of the divergence suggests that confrontation with the West is associated with a sustained elevation in perceived political risk, with implications for policy credibility, social coordination, and long-horizon economic decision-making.

In column (3), the estimated average post-confrontation effect on the rule-of-law index is -0.62 standard deviation, with an end-of-sample gap of approximately -1.06 standard deviation. In comparative perspective, a one-unit difference in rule-of-law measures often corresponds to the difference between countries in which contracts are generally enforceable and judicial decisions are relatively predictable, and countries in which legal outcomes are uncertain, politicized, or weakly enforced. Institutionally, this magnitude is comparable to the gap observed between countries such as Poland and Ukraine, or between Southern European states with relatively strong judicial enforcement (Spain, Slovenia) and those where legal predictability is substantially weaker (Greece, Albania). The estimated divergence therefore implies a shift away from rule-based governance toward a regime in which discretionary authority plays a more prominent role in economic and political interactions. The gradual widening of the gap over time is consistent with a process of institutional erosion, rather than an immediate breakdown of formal legal structures.

For control of corruption, the estimated average post-confrontation gap, reported in column (4), is -0.46 standard deviation, widening to approximately -0.71 standard deviation by the end of the sample. Differences of this magnitude typically distinguish countries in which corruption is episodic and constrained from those in which it is systemic and embedded in routine economic transactions. In practical terms, this gap corresponds to an environment in which access to scarce resources, such as foreign exchange, import licenses, procurement contracts, or regulatory approvals, is increasingly mediated by informal arrangements and political connections. The relatively rapid emergence of the corruption gap following the confrontation shock, combined with its persistence thereafter, suggests that confrontation induces durable rent-creation mechanisms rather than temporary governance slippage.

A central feature of the institutional results is their joint coherence across dimensions. The confrontation shock is associated with (i) declines of roughly one unit in political stability and rule of law, (ii) declines of approximately half to three-quarters of a unit in corruption control, and (iii) a multi-standard-deviation collapse in aggregate institutional capacity.

These magnitudes are comparable to those separating distinct institutional regimes, rather than marginal policy differences within a given regime. Just as a persistent GDP gap of 20-25 percent distinguishes countries on sharply different development paths, the estimated institutional gaps distinguish states operating under fundamentally different governance equilibria. Importantly, the institutional divergence mirrors the economic divergence in both timing and persistence. Institutions do not collapse instantaneously, nor do they recover quickly. Instead, they deteriorate gradually and cumulatively, consistent with a transition to a lower institutional equilibrium that is difficult to reverse.

*4.2.2.1    Summary and interpretation*

The magnitude of the institutional gaps provides a natural explanation for the scale and persistence of the economic losses documented earlier. A state characterized by lower political stability, weaker rule of law, and higher corruption is one in which investment horizons shorten, risk premia increase, and productive activity is increasingly displaced by rent-seeking and informality. The estimated institutional divergences are therefore quantitatively consistent with the observed collapse in foreign direct investment, reduced trade integration, and long-run output losses. In this sense, confrontation with the West operates not only through direct economic channels, but also through endogenous institutional deterioration, which amplifies and prolongs its economic effects. In sum, the institutional effects of Iran's confrontation with the West are large enough to be interpreted as real and persistent losses in state capacity and governance quality, comparable in magnitude to the institutional distances observed between countries on radically different development trajectories. The confrontation shock is associated with a sustained shift away from a rule-based, predictable institutional environment toward one characterized by instability, discretion, and rent allocation. These institutional losses provide a central mechanism through which the economic consequences of confrontation become long-lasting and difficult to reverse.

**Figure 2**: Institutional Trajectories under Iran's Confrontation with the West: Actual vs. Synthetic Control

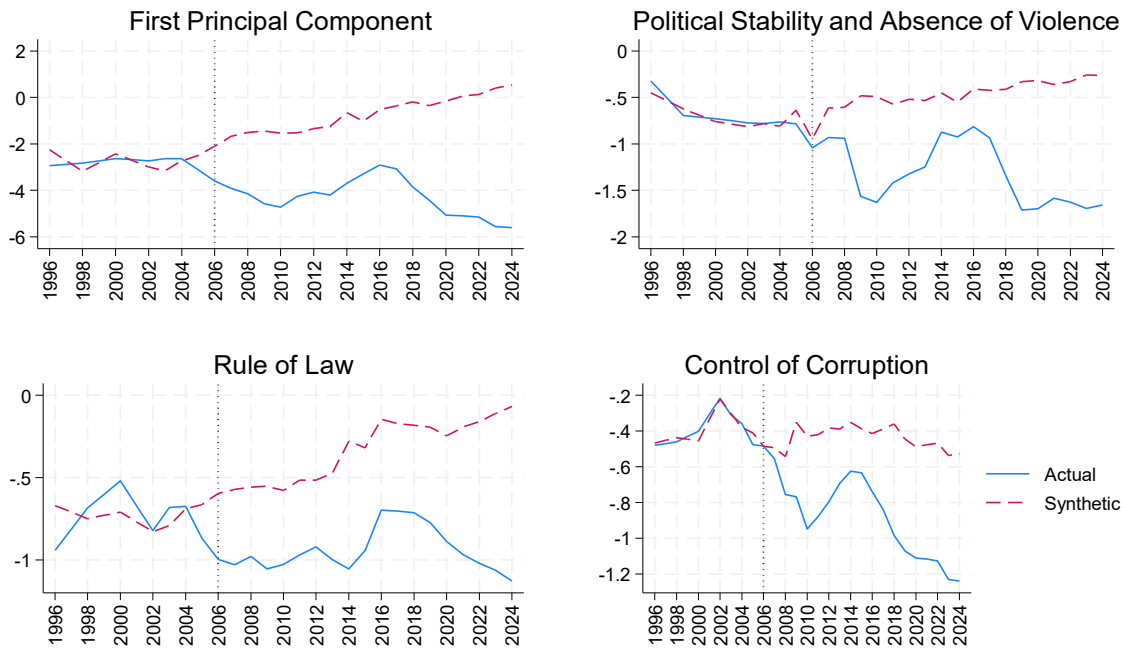

*4.3    In-space placebo analysis: permutation-based inference*

To assess whether the estimated post-confrontation divergences for Iran are unusually large relative to those that arise spuriously in the donor pool, we conduct an in-space placebo analysis based on permutation of the treatment assignment. This procedure follows the standard inferential logic in the synthetic control literature and provides a non-parametric benchmark against which to evaluate the magnitude and persistence of the estimated effects.

*4.3.1 Design of the in-space placebo analysis*

The in-space placebo exercise proceeds as follows. For each country in the donor pool, we reassign the confrontation date as if that country had experienced the same shock as Iran, re-estimate the synthetic control using the remaining donor countries, and compute the resulting post-treatment gaps. This generates an empirical distribution of placebo effects that captures the range of divergences that can arise mechanically from the synthetic control procedure when no true treatment is present. Inference is based on two complementary statistics. First, we compute the post-to-pre-treatment root mean squared prediction error (RMSPE) ratio, which summarizes how sharply the fit deteriorates after the assigned treatment date relative to the pre-treatment period. Second, we evaluate left-tailed permutation p-values for the estimated treatment effects at two horizons: (i) immediately after the onset of confrontation (T0+1), and (ii) at the end of the sample. The latter is particularly informative in the present context, as the confrontation with the West is conceptualized as a persistent, long-run shock rather than a transitory disturbance. Throughout, lower p-values indicate that Iran's estimated divergence is unusually negative relative to the distribution of placebo gaps in the donor pool. Table 7 reports the in-space placebo estimates.

**Table 7**: Permutation-Based In-Space Placebo Tests: Iran's Confrontation with the West

|  | Post/Pre-T0 RMSE (p-value) | Left-tailed p-value on H0 | | Verdict |
|---|---|---|---|---|
|  |  | T0+1 | End-of-sample |  |
|  | (1) | (2) | (3) | (4) |
| Panel A: Macroeconomic outcomes | | | | |
| Real GDP | 0.153 | 0.615 | 0.076 | Permanent negative |
| Real GDP per capita | 0.307 | 0.416 | 0.384 | Negative, weak significance |
| Trade openness | 0.846 | 0.307 | 0.231 | Negative weak significance |
| Non-oil export share | 0.307 | 0.231 | 0.231 | Negative, weak significance |
| FDI inflows | 0.077 | 0.384 | 0.076 | Permanent negative |
| Inflation rate | 0.153 | 0.076 | 0.231 | Temporary negative |
| Foreign exchange rate | 0.231 | 0.231 | 0.076 | Permanent negative |
| Military expenditure | 0.615 | 0.153 | 0.692 | Temporary, negative |
| PPP | 0.441 | 0.153 | 0.076 | Permanent negative |
| Panel B: Institutional outcomes | | | | |
| Principal component | 0.238 | 0.076 | 0.076 | Permanent negative |
| Political stability and absence of violence | 0.076 | 0.153 | 0.069 | Permanent negative |
| Rule of law | 0.307 | 0.076 | 0.076 | Permanent negative |
| Control of corruption | 0.076 | 0.307 | 0.153 | Permanent negative |

The in-space placebo results for macroeconomic outcomes reveal a clear and economically interpretable pattern. For real GDP, Iran's post-confrontation divergence is systematically larger than those observed in most placebo assignments. While short-horizon significance is moderate, the end-of-sample permutation p-value places Iran in the lower tail of the placebo distribution, indicating that the long-run GDP shortfall is unlikely to be a generic artifact of the synthetic control procedure. This pattern is consistent with the baseline interpretation of confrontation as a shock whose effects accumulate gradually over time rather than materializing instantaneously.

A similar conclusion emerges for foreign direct investment (FDI) inflows. Both the post/pre-treatment RMSPE ratio and the end-of-sample permutation p-value indicate that Iran's collapse in FDI is unusually severe relative to the donor pool. Given that FDI is forward-looking and highly sensitive to geopolitical risk and market access, this result provides particularly clean evidence of a persistent external constraint induced by confrontation. The foreign exchange rate and purchasing power parity measures also exhibit strong long-run divergence. In both cases, Iran lies in the lower tail of the placebo distribution at the end of the sample, consistent with a permanent deterioration in external balance conditions and relative prices rather than a temporary episode of volatility.

By contrast, several real-sector margins, most notably GDP per capita, trade openness, and non-oil exports, exhibit weaker permutation significance. Importantly, these outcomes are slow-moving, subject to higher measurement noise, and influenced by demographic and compositional factors that tend to reduce the power of permutation tests. The fact that their estimated effects are consistently negative and economically coherent with the stronger results for aggregate GDP and FDI suggests attenuation due to limited statistical power rather than absence of an underlying effect. Finally, the placebo analysis indicates that the effects on inflation and military expenditure are less persistent. Inflation exhibits a negative divergence in the immediate post-treatment period that does not consistently place Iran in the extreme tail of the placebo distribution at longer horizons, while military expenditure shows no robust long-run divergence. These findings are informative, as they indicate that the confrontation shock does not generate mechanically persistent effects across all macroeconomic aggregates.

The in-space placebo results for institutional outcomes are notably stronger and more uniform than those for several macroeconomic variables. For the institutional principal component, Iran's post-confrontation divergence is consistently among the most negative in the donor pool. Both the RMSPE ratio and the end-of-sample permutation p-value indicate that the magnitude of the institutional deterioration is unlikely to arise by chance under placebo reassignment. This result corroborates the baseline finding of a large and persistent collapse in aggregate institutional capacity.

The same pattern holds for political stability and absence of violence and for rule of law. In both cases, Iran's estimated post-confrontation gap places it firmly in the lower tail of the placebo distribution at the end of the sample, indicating a permanent institutional divergence rather than a short-lived governance shock. For control of corruption, the permutation evidence is somewhat weaker but remains directionally consistent. Iran exhibits sustained deterioration relative to the synthetic counterfactual, and while the end-of-sample p-value is larger than for political stability or rule of law, the magnitude of the gap remains unusual relative to most placebo assignments. Given the noisier nature of corruption measures, this pattern is consistent with a genuine but more difficult-to-detect long-run effect.

Taken together, the institutional placebo results indicate that Iran's post-confrontation institutional trajectory is systematically worse than that of the vast majority of placebo-treated donor countries, particularly at longer horizons. This stands in contrast to the mixed short-run significance and reinforces the interpretation of confrontation as a shock with enduring institutional consequences. The distinction between permanent and temporary effects in the placebo analysis is conceptual rather than mechanical. An outcome is classified as exhibiting a permanent negative effect if Iran remains in the lower tail of the placebo distribution at the end of the sample, indicating sustained divergence relative to the donor pool. By contrast, outcomes classified as temporary display an initial divergence that attenuates over time and does not remain unusually large relative to placebo assignments at longer horizons. Under this interpretation, the in-space placebo analysis provides strong evidence of permanent negative effects for real GDP, FDI, exchange rate dynamics, PPP measures, and all core institutional outcomes, while suggesting more transitory or weakly identified effects for inflation, military expenditure, and certain trade margins.

Overall, the in-space placebo analysis supports the central interpretation of the paper. Iran's post-confrontation economic and institutional divergences, particularly for aggregate output, capital flows, external balance variables, and institutional quality, are unusually large and persistent relative to those observed under placebo treatment of donor countries. While some outcomes exhibit weaker permutation significance, the pattern of results is internally coherent and consistent with a confrontation shock that operates through long-run equilibrium channels rather than short-run disturbances. In this sense, the placebo evidence reinforces the conclusion that the estimated effects reflect genuine long-run consequences of confrontation with the West, rather than artifacts of the synthetic control methodology or idiosyncratic features of the donor pool.

### 4.4   Generalized synthetic control estimates

To assess the robustness of the baseline results and to characterize the dynamic evolution of treatment effects, I re-estimate the economic impacts of Iran's confrontation with the West using the generalized synthetic control (GSC) framework with interactive fixed effects, following Xu (2017). This approach allows for unobserved common factors with unit-specific loadings and yields a full dynamic treatment effect path with confidence intervals, thereby addressing uncertainty surrounding the average treatment effect estimates. The GSC estimator complements the baseline synthetic control analysis along two dimensions. First, by modelling outcomes as driven by latent common factors, it allows for richer unobserved heterogeneity than standard two-way fixed effects and mitigates concerns that post-treatment divergence may be driven by unaccounted global or regional shocks. Second, it delivers time-specific average treatment effects on the treated ($ATT_{t>T_0}$) with associated confidence intervals, making it possible to assess whether estimated effects are gradual, persistent, or transitory. Given that the confrontation with the West is conceptualized as a long-run structural shock, the dynamic ATT profiles are particularly informative.

#### 4.4.1   Economic magnitudes

The generalized synthetic control estimates allow a direct translation of the estimated treatment effects into economically meaningful magnitudes while accounting for uncertainty in the dynamic treatment paths. Because several outcomes are specified in logarithmic form, the estimated average treatment effects can be interpreted as proportional deviations of Iran's post-confrontation outcomes from their counterfactual trajectories implied by the interactive fixed effects model.

Table 8 reports generalized synthetic control estimated gaps in response to the confrontation with the West. For aggregate output, the estimated average treatment effect on log real GDP in column (1), is -0.140 log points, with a 95 percent confidence interval of [-0.234, -0.049]. This estimate implies that, on average, Iran's real GDP is approximately 13 percent lower than its counterfactual level following the confrontation with the West, with plausible values ranging from about 5 percent to 21 percent. The dynamic treatment path indicates that this divergence deepens over time rather than dissipating, and by the end of the sample the estimated gap reaches -0.183 log points, corresponding to an output shortfall of roughly 17 percent. These magnitudes indicate that the confrontation shock is associated with a persistent downward shift in Iran's long-run output path, rather than a temporary loss followed by recovery.

The corresponding estimates for living standards, reported in column (2) are at least as large. The average effect of the confrontation on log real GDP per capita is -0.191 log points, with a 95 percent confidence interval of [-0.265, -0.097]. In level terms, this implies an average post-confrontation income shortfall of approximately 17 percent per person, with plausible values ranging from about 9 percent to 23 percent. By the end of the sample, the estimated gap in GDP per capita reaches -0.274 log points, implying a decline of roughly 24 percent relative to the counterfactual path. The close alignment between aggregate GDP and per-capita GDP effects indicates that population dynamics do not drive these results; instead, the confrontation is associated with a substantial and persistent reduction in income generation per capita.

These proportional gaps can be translated into real economic losses using transparent accounting assumptions. If Iran's counterfactual GDP at market exchange rates would have been on the order between 350 and 400 billion U.S. dollars per year over the post-confrontation period, an average output shortfall of approximately 13 percent implies annual forgone production of roughly 45-55 billion dollars, rising to approximately 60-70 billion dollars per year by the end of the sample as the gap deepens. Aggregated over the 2007-2024 period, this corresponds to cumulative losses on the order of roughly one trillion dollars at market exchange rates under conservative assumptions. Expressed in purchasing-power-parity terms, the implied annual losses are substantially larger, reflecting forgone domestic absorption rather than external purchasing power. These calculations are accounting translations of the estimated proportional gaps rather than welfare measures, and they are reported to convey the scale of the long-run divergence rather than to provide precise estimates of social loss.

The generalized synthetic control estimates also shed light on the magnitude of changes in Iran's external integration and diversification. For non-oil exports, the estimated average treatment effect of -0.161 log points in column (4) implies that non-oil exports are approximately 15 percent lower than their counterfactual level on average, with a 95 percent confidence interval indicating plausible declines between roughly 11 and 18 percent. The dynamic treatment path shows that this gap widens gradually over time, reaching approximately 16-17 percent by the end of the sample. These magnitudes suggest that confrontation with the West substantially constrained Iran's ability to diversify away from oil exports, consistent with long-run impediments to export learning, access to foreign inputs, and participation in global value chains. By contrast, the estimated effect on aggregate trade openness is smaller and imprecisely estimated in the generalized synthetic control framework, reflecting both slower adjustment on this margin and greater heterogeneity across countries.

For foreign direct investment inflows, which are measured in levels rather than logs, the estimated average treatment effect in column (5) is large and negative, indicating a substantial and persistent collapse in FDI relative to the counterfactual. Although the confidence intervals are wide due to the volatility of FDI flows, the dynamic treatment path shows an immediate and durable downward shift following the confrontation, with no evidence of recovery over time. This pattern reinforces the interpretation that confrontation induced a lasting increase in perceived country risk and reduced expectations of long-run market access, rather than a transitory investment response.

The magnitude of effects is also economically substantial. For the foreign exchange rate, which is specified in log-normalized form, the estimated average treatment effect in column (7) of 0.913 implies that the exchange rate index is approximately 2.5 times higher than its counterfactual level on average, with the end-of-sample effect exceeding 1.6 log points, corresponding to a more than five-fold increase relative to the counterfactual. These magnitudes indicate a persistent and cumulative depreciation relative to the path predicted in the absence of confrontation. In contrast, inflation exhibits greater volatility and weaker average effects, suggesting that the primary long-run nominal adjustment operates through the exchange rate rather than sustained inflation differentials. The estimated effect on the purchasing power parity measure in column (9) is likewise large and statistically significant, with an average log-normalized effect of 1.144, corresponding to a multiplicative increase of roughly three relative to the counterfactual and substantially larger effects by the end of the sample. Because PPP measures are mechanically linked to inflation and exchange-rate dynamics, these magnitudes should be interpreted as evidence of large and persistent relative-price distortions rather than as independent indicators of welfare changes.

Taken together, the magnitude analysis under the generalized synthetic control framework reinforces three central conclusions. First, confrontation with the West is associated with large and persistent losses in real output and living standards, on the order between 10 and 20 percent relative to a plausible counterfactual path. Second, these losses are accompanied by economically meaningful contractions in non-oil exports and foreign direct investment, indicating long-run constraints on diversification and capital inflows. Third, real adjustments, particularly through the exchange rate, are large and cumulative, consistent with a sustained external constraint rather than short-run macroeconomic instability. Importantly, the generalized synthetic control estimates attenuate the baseline synthetic control magnitudes but do not overturn them, strengthening the interpretation that the economic effects of confrontation reflect a durable shift in Iran's long-run equilibrium rather than transitory fluctuations. Figure 3 presents the generalized synthetic control estimates in greater detail.

**Figure 3**: Generalized Synthetic Control Estimates of the Economic Effects of Iran's Confrontation with the West, 1996-2024

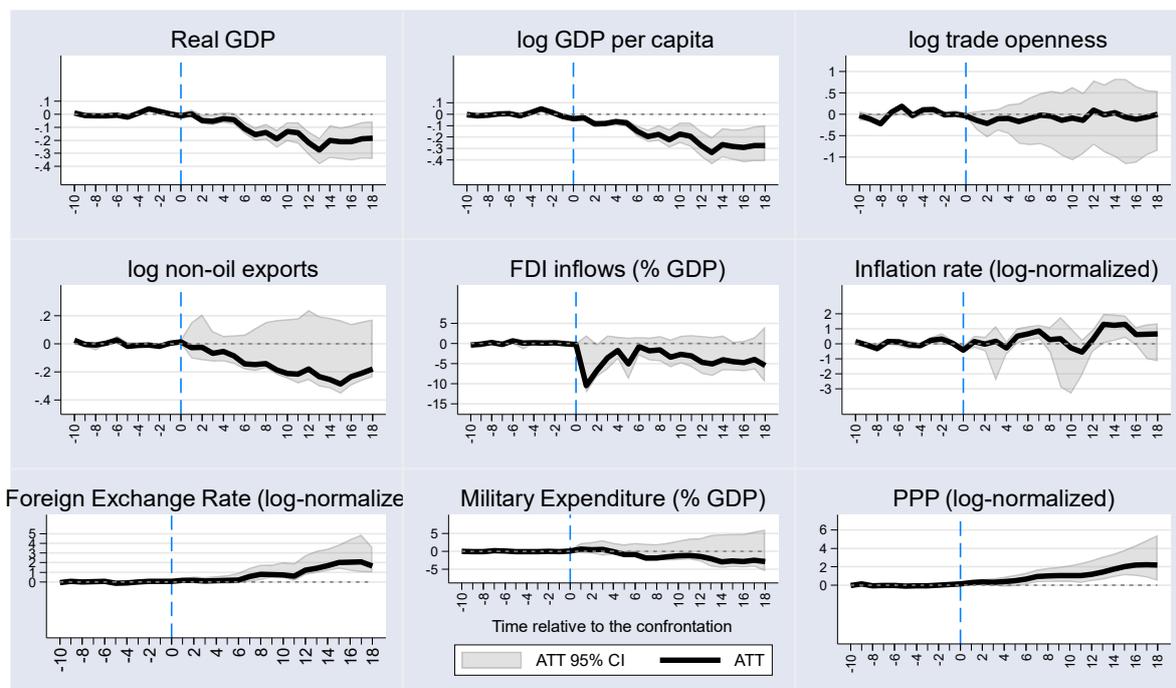

Table 8: Generalized Synthetic Control Estimates of the Economic Effects of Iran's Confrontation with the West

| | Real GDP (log) | Real GDP per capita (log) | Trade Openness (log) | Non-Oil Exports (log) | FDI net inflows, % | Inflation rate (log-normalized) | Foreign Exchange Rate (log-normalized) | Military Expenditure (% GDP) | Purchasing Power Parity (PPP) |
|---|---|---|---|---|---|---|---|---|---|
| | (1) | (2) | (3) | (4) | (5) | (6) | (7) | (8) | (9) |
| Average Effect | -0.140*** | -0.191*** | -0.077 | -0.161* | -4.086* | 0.433 | 0.913** | -1.382 | 1.144*** |
| | (0.043) | (0.044) | (0.350) | (0.087) | (2.437) | (0.398) | (0.377) | (1.417) | (0.460) |
| 95% Confidence Intervals | (-0.234,-0.049) | (-0.265, -0.097) | (-0.768, 0.442)) | (-0.199, 0.113) | (-6.411, 0.856) | (-0.734, 1.011) | (0.598, 1.971)) | (-2.304, 3.247)) | (0.598, 2.361) |
| End-of-sample effect | -0.183** | -0.274*** | 0.001 | -0.180 | -5.462 | 0.662 | 1.659 | -2.868 | 2.181 |
| | (0.084) | (0.085) | (0.414) | (0.109) | (3.722) | (0.568) | (0.679) | (2.872) | (1.220) |
| ATE p-value | 0.000 | 0.000 | 0.821 | 0.098 | 0.142 | 0.277 | 0.016 | 0.333 | 0.013 |
| In-time placebo test (p-value) | 0.500 | 0.362 | 0.171 | 0.994 | 0.162 | 0.798 | 0.111 | 0.001 | 0.112 |

Notes: This table reports generalized synthetic control estimates of the economic effects of Iran's confrontation with the West, implemented using an interactive fixed effects model following Xu (2017). The treatment is dated to 2006/2007, corresponding to the onset of a sustained confrontation regime. Each column reports results for a separate outcome. Columns (1) through (4) report outcomes in logarithms; column (5) reports net foreign direct investment inflows as a percentage of GDP; columns (6) and (7) report log-normalized inflation and exchange-rate indices; column (8) reports military expenditure as a percentage of GDP; column (9) reports a log-normalized purchasing power parity (PPP) measure. The generalized synthetic control estimator models untreated potential outcomes as a function of unit fixed effects, time fixed effects, and a low-dimensional latent factor structure that captures unobserved common shocks with heterogeneous factor loadings. The average effect reports the mean post-treatment average treatment effect on the treated (ATT). The end-of-sample effect reports the estimated ATT in the final year of the sample. The 95% confidence intervals are constructed using nonparametric bootstrap procedures appropriate for interactive fixed effects models. ATE p-values correspond to tests of the null hypothesis of no average post-treatment effect. The in-time placebo test reports p-values from reassigning the treatment to pre-treatment periods to assess potential anticipation effects. Asterisks denote statistically significant ATT parameters at 10% (*), 5% (**), and 1% (***), respectively.

*4.4.2 Institutional magnitudes*

The generalized synthetic control estimates for institutional outcomes provide a disciplined way to translate the estimated treatment effects into interpretable changes in governance quality, while accounting for latent common factors and time-varying heterogeneity. Because the institutional indicators are measured on standardized or bounded scales rather than in logs, magnitude interpretation necessarily proceeds through comparative benchmarks in the cross-country distribution of governance outcomes, rather than through direct percentage translations.

Table 9 reports the GSC estimated institutional magnitudes of Iran's confrontation with the West. We begin with the aggregate measure of institutional capacity, constructed as the first principal component of the underlying governance indicators. In column (1), the estimated average treatment effect is -1.71, with a 95 percent confidence interval of [-2.51, -0.96], and the end-of-sample effect reaches -3.05. In comparative terms, a one-unit change in a standardized institutional principal component typically corresponds to a substantial repositioning within the global distribution of state capacity. An average decline of approximately 1.7 units therefore implies a large downward shift in Iran's institutional rank relative to its counterfactual, while the end-of-sample gap exceeding three units places Iran far outside the range associated with incremental institutional reform or cyclical political variation. The dynamic treatment path shows that this deterioration unfolds gradually after the confrontation with the West and persists throughout the sample period, consistent with cumulative institutional erosion rather than a transitory governance shock. Figure 4 presents the estimates at greater depth.

**Table 9**: Generalized Synthetic Control Estimates of the Institutional Effects of Iran's Confrontation with the West

|  | Principal Component | Political Stability | Rule of Law | Control of Corruption |
|---|---|---|---|---|
|  | (1) | (2) | (3) | (4) |
| Average Effect | -1.711*** | -0.839** | -0.275*** | -0.506*** |
|  | (0.415) | (0.387) | (0.088) | (0.074) |
| 95% Confidence Intervals | (-2.511, -0.958) | (-1.539, 0.221) | (-0.425, -0.091) | (-0.663, -0.345) |
| End-of-sample effect | -3.051*** | -1.301*** | -0.591*** | -0.809*** |
|  | (0.611) | (0.483) | (0.118) | (0.133) |
| ATE p-value | 0.000 | 0.003 | 0.002 | 0.000 |

Notes: This table reports generalized synthetic control estimates of the institutional effects of Iran's confrontation with the West, implemented using an interactive fixed effects model following Xu (2017). The treatment is dated to 2006/2007, corresponding to the onset of a sustained confrontation regime. Each column reports results for a separate institutional outcome drawn from the Worldwide Governance Indicators (WGI). Column (1) reports the first principal component of institutional quality; columns (2) through (4) report Political Stability and Absence of Violence, Rule of Law, and Control of Corruption effects, respectively. Outcomes are standardized following the WGI methodology, with higher values indicating stronger institutional performance. The generalized synthetic control estimator models untreated potential outcomes using unit fixed effects, time fixed effects, and a low-dimensional latent factor structure that captures unobserved common shocks with heterogeneous factor loadings. The Average Effect reports the mean post-treatment average treatment effect on the treated (ATT). The End-of-sample effect reports the estimated ATT in the final year of the sample. The 95% confidence intervals are constructed using nonparametric bootstrap procedures appropriate for interactive fixed effects models. ATE p-values correspond to tests of the null hypothesis of no average post-treatment effect. Asterisks denote statistically significant ATT parameters at 10% (*), 5% (**), and 1% (***), respectively.

For political stability and absence of violence, the estimated average treatment effect in column (2) is -0.84, with a 95 percent confidence interval of [-1.54, -0.22], and an end-of-sample gap of approximately -1.30. In the cross-country distribution of political stability indices, differences of this magnitude typically separate countries with relatively predictable political environments from those in which political violence, repression, or severe instability is a persistent risk. Interpreted in this way, the GSC estimates imply that confrontation with the West is associated with a qualitative deterioration in political predictability, rather than a marginal increase in episodic unrest. The dynamic path indicates that this deterioration is sustained over time, reinforcing the interpretation of confrontation as a long-run destabilizing force.

The estimates for rule of law in column (3) point in the same direction. The average treatment effect is -0.28, with a 95 percent confidence interval of [-0.43, -0.09], and the end-of-sample gap approaches -0.60. In comparative perspective, a decline of roughly half a unit in rule-of-law indices is commonly associated with a shift from environments in which contracts are generally enforceable and judicial decisions are predictable to environments characterized by greater discretion, politicization, and uncertainty in legal outcomes. The magnitude of the estimated gap therefore implies a meaningful weakening of legal constraints on both public and private actors, with direct implications for investment incentives and the allocation of economic activity. The gradual widening of the gap over time is consistent with a process of institutional erosion rather than abrupt legal breakdown.

For control of corruption, column (4) reports the estimated average treatment effect of -0.51 basis points, with a 95 percent confidence interval of [-0.66, -0.35], and the end-of-sample gap reaches approximately -0.81. Differences of this size typically distinguish countries where corruption is episodic and constrained from those where it is systemic and embedded in routine economic transactions. In practical terms, this magnitude implies a substantial increase in the role of informal payments, political connections, and rent allocation in mediating access to scarce resources such as foreign exchange, licenses, procurement contracts, or regulatory approvals. The persistence and monotonicity of the dynamic treatment path suggest that confrontation generates durable rent-creation mechanisms, rather than temporary governance slippage.

Taken together, the institutional GSC magnitudes indicate a coherent and economically meaningful shift in Iran's governance equilibrium following confrontation with the West. Average declines on the order of one unit in political stability and half to three-quarters of a unit in rule of law and corruption control, combined with multi-unit collapses in the aggregate institutional principal component, are comparable in scale to the institutional distances observed between countries on sharply different development trajectories. Just as a persistent GDP gap of 15-25 percent separates economies such as South Korea and Argentina or Poland and Ukraine, the estimated institutional gaps separate states operating under fundamentally different governance regimes.

Importantly, these institutional magnitudes align closely with the generalized synthetic control results for economic outcomes. A state characterized by lower political stability, weaker rule of law, and higher corruption is one in which investment horizons shorten, risk premia rise, and productive activity is increasingly displaced by rent-seeking and informality. The estimated institutional gaps are therefore quantitatively consistent with the observed long-run declines in output, non-oil exports, and foreign direct investment documented earlier. In this sense, the institutional GSC results provide a mechanism-consistent bridge between confrontation and persistent economic loss.

Two caveats are worth emphasizing. First, the institutional magnitudes should be interpreted as relative deviations from a counterfactual governance trajectory, not as absolute measures of institutional "failure." Second, while standardized indices facilitate cross-country comparison, they do not capture all dimensions of governance quality or distributional effects within countries. Nonetheless, the size, persistence, and coherence of the estimated gaps leave little doubt that confrontation with the West is associated with large and lasting losses in institutional capacity, which in turn help explain why the economic effects of confrontation are so persistent and difficult to reverse.

**Figure 4**: Dynamic Institutional Effects of Iran's Confrontation with the West: Generalized Synthetic Control Estimates

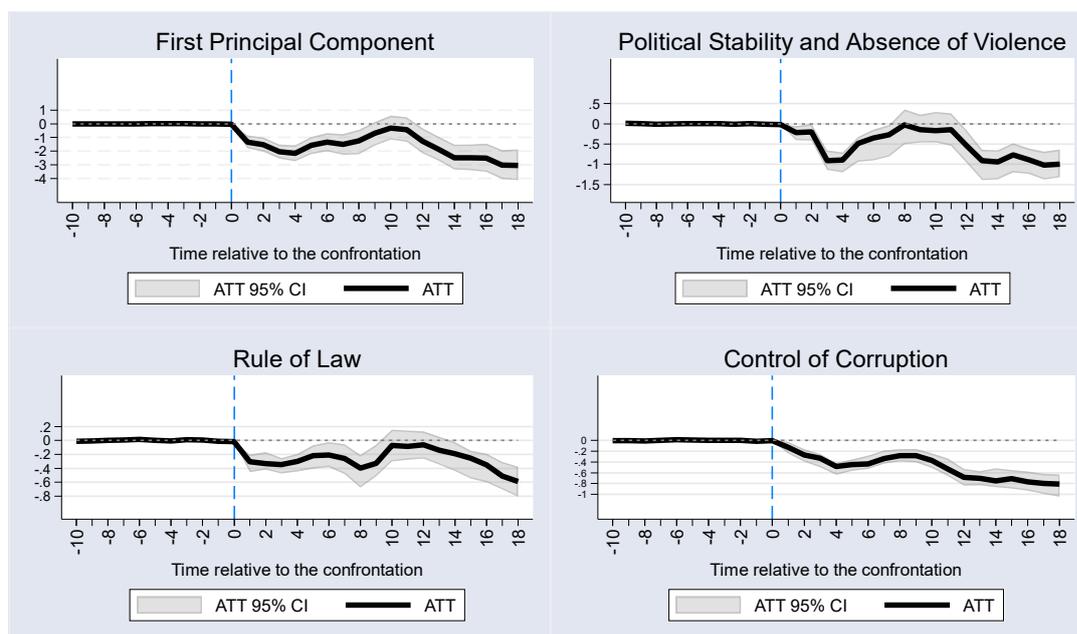

*4.5    Discussion and policy implications*

This paper studies the long-run economic and institutional consequences of Iran's confrontation with the West using a combination of synthetic control methods, generalized synthetic control with interactive fixed effects, and multiple layers of placebo inference. Taken together, the results point to a coherent and robust conclusion: confrontation operates as a persistent economic and institutional shock, inducing a durable downward shift in both economic performance and governance quality that is large by international standards and difficult to reverse.

The magnitude and persistence of the estimated output losses place the findings squarely within the literature documenting non-reverting growth paths after major disruptions. Seminal work by Cerra and Saxena (2008) shows that large negative shocks such as financial crises or wars, typically generate permanent output losses rather than temporary deviations from trend. Our results extend this insight to the context of sustained geopolitical confrontation in peacetime, demonstrating that confrontation alone can induce long-run GDP and GDP-per-capita gaps of comparable magnitude. The estimated output losses are also consistent with a broader body of evidence emphasizing path dependence in economic development following large shocks (Davis and Weinstein, 2002, Brakman et. al. 2004, Bleakley and Lin 2012). In contrast to these studies, which focus primarily on physical destruction or spatial reallocation, our analysis highlights a channel operating through external isolation and institutional erosion, even in the absence of widespread domestic violence. Methodologically, the use of synthetic control methods follows the framework developed by Abadie and Gardeazabal (2003) and Abadie et. al. (2015), who show that SCM is particularly well suited for studying the long-run effects of large, non-recurrent shocks (Gilchrist et. al. 2023). The persistence of the estimated gaps, combined with strong in-space and in-time placebo performance, supports interpreting the post-confrontation divergence as a genuine shift in Iran's growth trajectory rather than an artifact of global trends or transitory disturbances.

The central contribution lies in documenting that confrontation affects not only economic outcomes but also core dimensions of institutional quality. This finding complements the large literature emphasizing institutions as fundamental determinants of long-run development (Acemoglu et. al. 2001). While much of this work treats institutional differences as historically determined and slow-moving, our results indicate that sustained external confrontation can itself generate substantial institutional divergence over time. The estimated declines in political stability, rule of law, and corruption control are economically large and persistent, with magnitudes comparable to those separating distinct governance regimes in cross-country data. This pattern aligns closely with the framework developed by Besley and Persson (2011), in which conflict and external threats weaken state capacity by redirecting incentives toward short-term survival and rent extraction rather than long-run institutional investment. It is also consistent with the notion of a transition toward a limited-access order, as described by North et. al. (2009), in which scarcity and insecurity increase the role of discretion and rents in mediating economic activity. Importantly, the institutional deterioration documented here is gradual rather than abrupt, suggesting a process of cumulative erosion rather than sudden breakdown. This dynamic is consistent with evidence from conflict and violence settings showing that institutional damage often persists long after the initial shock has passed (Dell and Querubin 2018, Besley and Reynal-Querol 2014).

The scale of the estimated economic and institutional losses invites direct comparison with the literature on the costs of armed conflict and civil war. Abadie and Gardeazabal (2003) and Collier et al. (2004) document GDP losses on the order between 20 percent and 30 percent in regions or countries affected by internal armed conflict. Strikingly, the magnitudes estimated in this paper, both for output and for institutional capacity, are of similar order, despite the absence of large-scale domestic warfare. This comparison suggests that sustained geopolitical confrontation can generate civil-war-scale economic and institutional losses in peacetime. The key difference lies not in the magnitude of the losses, but in the mechanism: rather than physical destruction or population displacement, confrontation operates through prolonged isolation, elevated risk premia, and endogenous institutional decay.

The collapse of foreign direct investment and the persistence of exchange-rate distortions documented in the analysis further connect the findings to the literature on risk, capital flows, and irreversibility. Existing work emphasizes that weak institutions and heightened political risk substantially reduce capital inflows (Alfaro et al. 2008) and that misallocation induced by external constraints can have long-lasting growth effects (Gourinchas and Jeanne 2013). The persistence of the FDI collapse in our results is consistent with the idea that once investor expectations adjust downward, capital inflows may not recover even if external conditions partially improve. While a substantial literature examines the economic effects of sanctions (Hufbauer et al., 2007, Neuenkirch and Neumeier, 2015, Gutmann et. al. 2023), our findings suggest that sanctions should be understood as one component of a broader confrontation regime whose long-run effects extend well beyond the direct trade and financial channels typically studied. In particular, the institutional erosion documented here implies that the economic costs of confrontation may persist long after specific sanctions are lifted.

The results carry several important policy implications. First, they suggest that the economic costs of sustained confrontation are structural rather than transitory. From a policy perspective, confrontation should therefore be viewed not as a reversible strategy with short-run costs, but as a choice that reshapes long-run economic and institutional equilibria. Second, the findings highlight the central role of institutions as a transmission mechanism. Policies aimed solely at restoring trade or financial access, without addressing the institutional damage associated with prolonged isolation, are unlikely to generate a full economic recovery. Once political stability, legal predictability, and corruption control deteriorate to the extent documented here, rebuilding investor confidence and productive capacity becomes substantially more difficult. Third, the asymmetry between economic and institutional adjustment implies that institutional recovery may lag economic normalization by many years. Even if confrontation is partially relaxed, the legacy of weakened institutions may continue to depress growth, investment, and diversification. This insight is relevant not only for Iran but also for other countries experiencing prolonged periods of geopolitical isolation.

Finally, the findings have broader implications for the use of confrontation as a tool of international pressure. The evidence suggests that sustained confrontation can impose civil-war-scale economic and institutional costs in peacetime, raising important questions about the long-run consequences of such strategies for both target countries and the international system. Taken together, the evidence supports a clear conclusion. Iran's confrontation with the West has imposed large, persistent, and mutually reinforcing economic and institutional costs, comparable in magnitude to those associated with major internal conflicts. By documenting these effects rigorously and situating them within the top-tier literature on growth persistence, institutions, and conflict, the paper contributes to a deeper understanding of how geopolitical confrontation can reshape long-run development trajectories through both economic and institutional channels.

## 5  Conclusion

This paper examines the long-run economic and institutional consequences of Iran's confrontation with the West. Using synthetic control methods, generalized synthetic control with interactive fixed effects, and extensive placebo inference, we construct credible counterfactual trajectories for economic performance and institutional quality over nearly two decades. Across methods and outcomes, a consistent pattern emerges: sustained geopolitical confrontation operates as a joint economic and institutional equilibrium shock, generating large, persistent losses that do not dissipate over time.

On the economic dimension, confrontation is associated with a durable downward shift in Iran's growth path. Real GDP and GDP per capita fall substantially below their counterfactual trajectories, foreign direct investment collapses, non-oil exports contract, and exchange-rate distortions accumulate. These effects emerge gradually and persist, indicating a structural divergence rather than a temporary disruption. While generalized synthetic control estimates attenuate the magnitudes relative to baseline synthetic control results, the direction, persistence, and economic significance of the effects remain unchanged. This pattern strengthens, rather than weakens, the interpretation that confrontation alters long-run economic equilibria.

On the institutional dimension, the evidence is equally decisive. Political stability deteriorates, rule of law weakens, corruption intensifies, and aggregate institutional capacity declines sharply relative to a credible counterfactual. The estimated magnitudes are comparable to those separating distinct governance regimes in the cross-country distribution. Importantly, institutional deterioration unfolds cumulatively and persists throughout the sample period, suggesting that institutions are not merely slow-moving background conditions but endogenous outcomes of prolonged external confrontation. Taken together, the economic and institutional results reveal a self-reinforcing mechanism. External confrontation raises risk premia and restricts access to markets and capital, reducing investment and growth. Scarcity and uncertainty, in turn, expand rents and weaken institutional constraints, further discouraging productive activity. Once this feedback loop is established, recovery becomes difficult even if external pressures are partially relaxed. In this sense, confrontation reshapes not only outcomes but the rules under which economic and political interactions take place.

A central implication of the analysis is that the magnitude of the losses documented here is comparable to those associated with civil wars and revolutionary episodes, despite the absence of large-scale domestic violence. Persistent output losses on the order of between 15 and 25 percent and institutional declines of similar scale have historically been linked to internal conflict. The Iranian case demonstrates that peacetime geopolitical confrontation can generate civil-war-scale economic and institutional costs, operating through isolation, elevated risk, and endogenous institutional decay rather than physical destruction. The findings have broader relevance beyond Iran. They suggest that sustained confrontation should not be viewed as a reversible policy stance with primarily short-run economic costs. Instead, confrontation constitutes a structural choice that can permanently alter development trajectories by jointly shifting economic incentives and institutional equilibria. Policies that focus narrowly on restoring trade or financial access, without addressing the institutional legacy of prolonged isolation, are unlikely to deliver full recovery once institutional erosion has taken hold.

More broadly, this paper contributes to a growing literature emphasizing that long-run development paths are shaped not only by internal institutions or violent conflict, but also by persistent external political environments. Geopolitical confrontation, when sustained, can reorder domestic economic and institutional equilibria in ways that are both deep and lasting. Understanding these dynamics is essential for evaluating the long-run costs of confrontation in an increasingly fragmented global economy.